\pdfoutput=1
\documentclass[fleqn,usenatbib]{mnras}

\usepackage{newtxtext,newtxmath}
\usepackage{tabularx}
\usepackage{gensymb}
\usepackage{hyperref}

\usepackage[T1]{fontenc}
\usepackage{ae,aecompl}

\usepackage{graphicx}	
\usepackage{amsmath}	
\usepackage{amssymb}	

\newcommand{\Gaia}{\textit{Gaia}}
\newcommand{\kepler}{\textit{Kepler}}
\newcommand{\corot}{\textit{CoRoT}}
\newcommand{\Kepler}{\textit{Kepler}}
\newcommand{\TESS}{\textit{TESS}}

\newcommand{\XMM}{\textit{XMM-Newton}}

\newcommand{\NGTS}{NGTS}
\newcommand{\teff}{$T_{\rm eff}$}

\newcommand{\longobjname}{NGTS J121939.5-355557}
\newcommand{\objname}{NGTS J1219-3555}
\newcommand{\longobjnametwomass}{2MASS J12193970-3556017}

\newcommand{\oscfrac}{$\frac{\Delta F_{osc}}{F_{tr}}$}

\newcommand{\flareenergy}{$3.2\pm^{0.4}_{0.3}\times 10^{36}$erg}
\newcommand{\energyone}{$2.3\pm0.2\times 10^{34}$erg}
\newcommand{\energytwo}{$7.2^{+0.7}_{-0.6}\times 10^{33}$erg}
\newcommand{\energythree}{$2.9\pm0.3\times 10^{33}$erg}
\newcommand{\flareenergynoerg}{$3.2\pm^{0.4}_{0.3}\times 10^{36}$}
\newcommand{\maxampvalue}{$7.2\pm0.8$}

\newcommand{\gaiabprp}{$\text{G}_{\text{BP}}-\text{G}_{\text{RP}}$}
\newcommand{\countrate}{$175\pm20$}
\newcommand{\Alfven}{Alfv\'en}
\newcommand{\Msun}{$\textrm{M}_\odot$}
\newcommand{\Rsun}{$\textrm{R}_\odot$}
\usepackage{siunitx}
\newcommand{\jj}[1]{{#1}}

\title[QPPs on an M star superflare]{Detection of a giant flare displaying quasi-periodic pulsations from a pre-main sequence M star with NGTS}

\author[J. A. G. Jackman et al.]{
James A. G. Jackman,$^{1,2}$\thanks{E-mail: J.Jackman@warwick.ac.uk}
Peter J. Wheatley,$^{1,2}$\thanks{E-mail: P.J.Wheatley@warwick.ac.uk}
Chloe E. Pugh,$^{1,2}$\newauthor
Dmitrii Y. Kolotkov$^{1,2}$,
Anne-Marie Broomhall,$^{1,2,3}$
Grant~M.~Kennedy,$^{1,2}$\newauthor
Simon J. Murphy,$^{4}$
Roberto~Raddi,$^{5,1,2}$
Matthew~R.~Burleigh,$^{6}$
Sarah~L.~Casewell,$^{6}$\newauthor
Philipp Eigm\"uller,$^{7,8}$
Edward Gillen,$^{9}$
Maximilian~N.~G{\"u}nther,$^{9}$
James S. Jenkins,$^{10}$\newauthor
Tom Louden,$^{1,2}$
James McCormac,$^{1,2}$
Liam Raynard,$^{6}$
Katja Poppenhaeger,$^{11}$\newauthor
St\'{e}phane Udry,$^{12}$
Christopher A. Watson,$^{11}$
Richard G. West,$^{1,2}$
\\
$^{1}$Dept. of Physics, University of Warwick, Gibbet Hill Road, Coventry CV4 7AL, UK \\
$^{2}$Centre for Exoplanets and Habitability, University of Warwick, Gibbet Hill Road, Coventry CV4 7AL, UK\\
$^{3}$Institute of Advanced Studies, University of Warwick, Coventry CV4 7HS, UK\\
$^{4}$School of Physical, Environmental and Mathematical Sciences, University of New South Wales Canberra, ACT 2612, Australia\\
$^{5}$Dr. Remeis-Sternwarte, Friedrich Alexander Universit\"{a}t Erlangen-N\"{u}rnberg,  Sternwartstr. 7, 96049  Bamberg, Germany\\
$^{6}$Dept.\ of Physics and Astronomy, Leicester Institute of Space and Earth Observation, University of Leicester, University Road,\\ Leicester, LE1 7RH, UK\\
$^{7}$Institute of Planetary Research, German Aerospace Center, Rutherfordstrasse 2, 12489 Berlin, Germany\\
$^{8}$Center for Astronomy and Astrophysics, TU Berlin, Hardenbergstr. 36, D-10623 Berlin, Germany\\
$^{9}$Astrophysics Group, Cavendish Laboratory, J.J. Thomson Avenue, Cambridge CB3 0HE, UK\\
$^{10}$Departamento de Astronomia, Universidad de Chile, Casilla 36-D, Santiago, Chile\\
$^{11}$Astrophysics Research Centre, Queen's University of Belfast, 1 University Road, Belfast BT7 1NN, UK \\
$^{12}$Department of Astronomy of the University of Geneva, Observatoire de Gen\'{e}ve, 51 ch des Maillettes, 1290 Versoix, Switzerland\\
}

\date{Accepted XXX. Received YYY; in original form ZZZ}

\pubyear{2015}

\begin{document}
\label{firstpage}
\pagerange{\pageref{firstpage}--\pageref{lastpage}}
\maketitle

\begin{abstract}
We present the detection of an energetic flare on the pre-main sequence M3 star \longobjname, which we estimate as only 2 Myr old. The flare had an energy of \flareenergy\ and a fractional amplitude of \maxampvalue, making it one of the most energetic flares seen on an M star. The star is also X-ray active, in the saturated regime with $log L_{X}/L_{Bol} = -3.1$. 
In the flare peak we have identified multi-mode quasi-periodic pulsations formed of two statistically significant periods of approximately 320 and 660 seconds. This flare is one of the largest amplitude events to exhibit such pulsations.
The shorter period mode is observed to start after a short-lived spike in flux lasting around 30 seconds, which would not have been resolved in \kepler\ or \TESS\ short cadence modes. 
Our data shows how the high cadence of \NGTS\ can be used to apply solar techniques to stellar flares and identify potential causes of the observed oscillations. We also discuss the implications of this flare for the habitability of planets around M star hosts and how NGTS can aid in our understanding of this.
\end{abstract}




\section{Introduction}
Stellar flares from M stars provide some of the most dramatic stellar events, yet they cannot be predicted beforehand. Catching the relatively-rare high-energy events therefore requires long-duration measurements of many stars, such as that from wide-field surveys for transiting exoplanets. 
The energies output in such events often dwarf the largest seen from the Sun \citep[$10^{32}$ erg for the Carrington Event,][]{Carrington_1859,Carrington_Energy} and the flares themselves can be seen over a wide range of wavelengths, notably in the optical, ultraviolet and in X-rays. The high energy irradiation provided by these flares, combined with the discoveries of habitable zone exoplanetary systems, has in recent years provoked discussion of the role of flares in habitability. Along with a range of detrimental effects such as ozone depletion \citep[][]{Segura10} and potential atmospheric loss \citep[e.g.][]{Lammer07}, stellar flares have been invoked as a possible way of providing the NUV flux required for prebiotic chemistry on M dwarf exoplanetary systems \citep[e.g.][]{Ranjan17,Rimmer18}. Determining the role of stellar flares in exoplanetary habitability requires observations of not only more common lower energy events but also the rarest high energy ones, along with tracking how the occurrence of such events varies with stellar age. 

When observed in the optical, the highest energy white-light flares from mid-M and later spectral types can change the brightness of the star by magnitudes for hours before returning to quiescence \citep[e.g.][]{Gizis17,Paudel18}. The largest amplitude events can also make stars that normally reside below the sensitivity of a survey visible for short periods of time \citep[e.g. the $\Delta\,V<-11$ flare from ASASSN-16ae identified by][]{Schmidt16}. These flares provide evidence for strong magnetic activity being present on stars well past the fully-convective boundary of M3-M4. This boundary marks where stellar interiors no longer contain a radiative zone and magnetic fields are generated by an alternative dynamo mechanism to earlier spectral types \citep[e.g.][]{Houdebine15,Houdebine17}. While rare for individual stars, surveys with large fields-of-view can make regular detections of these high amplitude events.
In recent years such flares have even been identified as a potential ``fog'' for future large-scale surveys such as LSST, due to their ability to mimic other astrophysical transient events \citep[e.g.][]{Kulkarni06,Berger12,Berger13}. 

Some large amplitude stellar flares have exhibited complex substructure. One type of substructure is the oscillations of flare intensity with time, commonly referred to as quasi-periodic pulsations (QPPs). A common phenomenon on the Sun \citep{2010SoPh..267..329K, 2015SoPh..290.3625S, 2016ApJ...833..284I}, QPPs remain relatively rare in stellar flare observations. Those which have been observed have been seen in the optical \citep[e.g.][]{Balona15}, microwave \citep[][]{Zaitsev04}, UV \citep[][]{Welsh06,Doyle18} and X-ray \citep[][]{Mitra05, Pandey09,Cho16}.

The exact cause of these pulsations is not yet known, however numerous mechanisms have been proposed and it is indeed possible that different mechanisms act in different cases. These mechanisms can be split into two groups: those where the flare emission is modulated by magnetohydrodynamic (MHD) oscillations, and those based on some regime of repetitive magnetic reconnection \citep[][]{Nakariakov16}. In the first case, MHD oscillations may directly affect the flare emission by modulating the plasma parameters \citep[e.g.][]{2009SSRv..149..119N}; for example variations of the magnetic field strength would cause gyrosynchrotron emission to vary. In addition the oscillations could indirectly affect the emission by modulating the kinematics of charged particles in the flaring coronal loops, which in turn would cause bremsstrahlung emission at the footpoints to appear periodically \citep[e.g.][]{2008PhyU...51.1123Z}. 
For the second case, the repetitive reconnection could be the result of external triggering, for example by leakage of MHD oscillations of a nearby structure \citep{Nakariakov06}. Alternatively it could result from a `magnetic dripping' mechanism which does not require an external trigger \citep[e.g.][]{2018SSRv..214...45M}. For this scenario, only when some threshold energy is reached does magnetic reconnection occur. This reconnection then releases energy and the process repeats periodically.
Examples of this regime are reported by \citet{2009A&A...494..329M, 2012A&A...548A..98M, 2017ApJ...844....2T}. If the mechanism behind a particular observation of QPPs in a flare can be determined, then it may be possible to estimate coronal plasma parameters in the vicinity of the flare via coronal seismology \citep[e.g.][]{2016SoPh..291.3143V}.

The timescale of QPPs can range from milliseconds to minutes \citep[e.g.][]{2018SSRv..214...45M} and as a result, without observations of a similar or higher cadence these short-lived behaviours will be missed. For solar flares this cadence is regularly achieved \citep[e.g.][]{Dolla12,Kumar17}, as it can be for targeted observations of stellar flares for individual stars \citep[e.g.][]{Mathioudakis03}. However, in order to identify more of these events on stellar flares, long duration observations with a high cadence are required for a large number of stars. This became possible with the \kepler\ satellite, which was launched in 2009 and observed over 195,000 stars \citep[][]{Huber14}. Indeed, short cadence observations from \Kepler\ have been used to increase the number of white-light QPP detections \citep[e.g.][]{Balona15,Pugh16}.
However, only a small fraction of targets were observed in the one-minute short cadence mode \citep[e.g.][]{Gilliland10} and not for the full duration of the mission, reducing the potential for detecting short timescale QPPs. Consequently it is apparent that to increase not only the number of both QPPs and high energy flares observed, long duration, wide field observations are required with a high cadence for all stars. 

This has become possible with the Next Generation Transit Survey (\NGTS). \NGTS\ is a ground-based transiting exoplanet survey and consists of 12 telescopes, each with a 520-890nm bandpass \citep[][]{Wheatley18}. 
Each camera operates with an exposure time of 10 seconds and has a field of view of $\approx$ 8 square degrees, resulting in a total instantaneous field of view of 96 square degrees. \NGTS\ also benefits from high-precision autoguiding \citep[][]{McCormac13} which, when combined with a pixel scale of 4.97 arcseconds per camera, enables the use of centroid analysis to rule out false positive planet candidates due to blended sources \citep[][]{Gunther17}. With stable tracking and wide fields, \NGTS\ is able to detect and resolve flares on both single and blended objects. 

In this paper we present the detection of a high energy stellar flare with QPPs from the pre-main sequence M star \longobjname. This is one of the best resolved observations of stellar flare substructure from a wide-field survey, allowing us to apply methods developed for Solar flares.
 We present our detection with \NGTS\ and discuss how we identified the source of the flare and derived its properties. We also present the oscillations of the flare and assess their significance, along with comparing them to previous observations of stellar and solar QPPs.

\section{Observations}
The data presented in this paper were collected with \NGTS\ over 115 nights between 2015 November 28th and 2016 August 4th. The detected flare occurred on the night of 2016 January 31st. 
\subsection{Flare Detection} \label{sec:flare_detect}
To identify flares in our data, we initially detrended the raw \NGTS\ lightcurves using a custom version of the {\scshape sysrem} algorithm. The full details of the \NGTS\ detrending pipeline can be found in \citet{Wheatley18}.

Stellar flares typically occur on timescales of minutes to hours, meaning most flares will have a duration less than a single night. We searched for flares on individual nights to make the most of this short timescale.

When searching for flares within a single night we employ a two-step flagging method. Initially we search for regions in the night where at least three consecutive data points are six median absolute deviations (MAD) above the night median \citep[as done by][]{Jackman18}. For the majority of flares, due to the aforementioned timescales, the median and MAD will not be strongly altered. This method therefore typically finds flares that occur purely within a single night. 
To find flares that dominate the entire night we also check whether the median of a night is five MAD above the median of the entire lightcurve. This is indicative of a longer timescale flare where we may only capture a portion of the event. After we have run this flagging procedure, we visually inspect all candidates to remove false positives. Examples of false positives include high amplitude variable stars and satellites passing through our aperture.

\begin{figure}
	\includegraphics[width=\columnwidth]{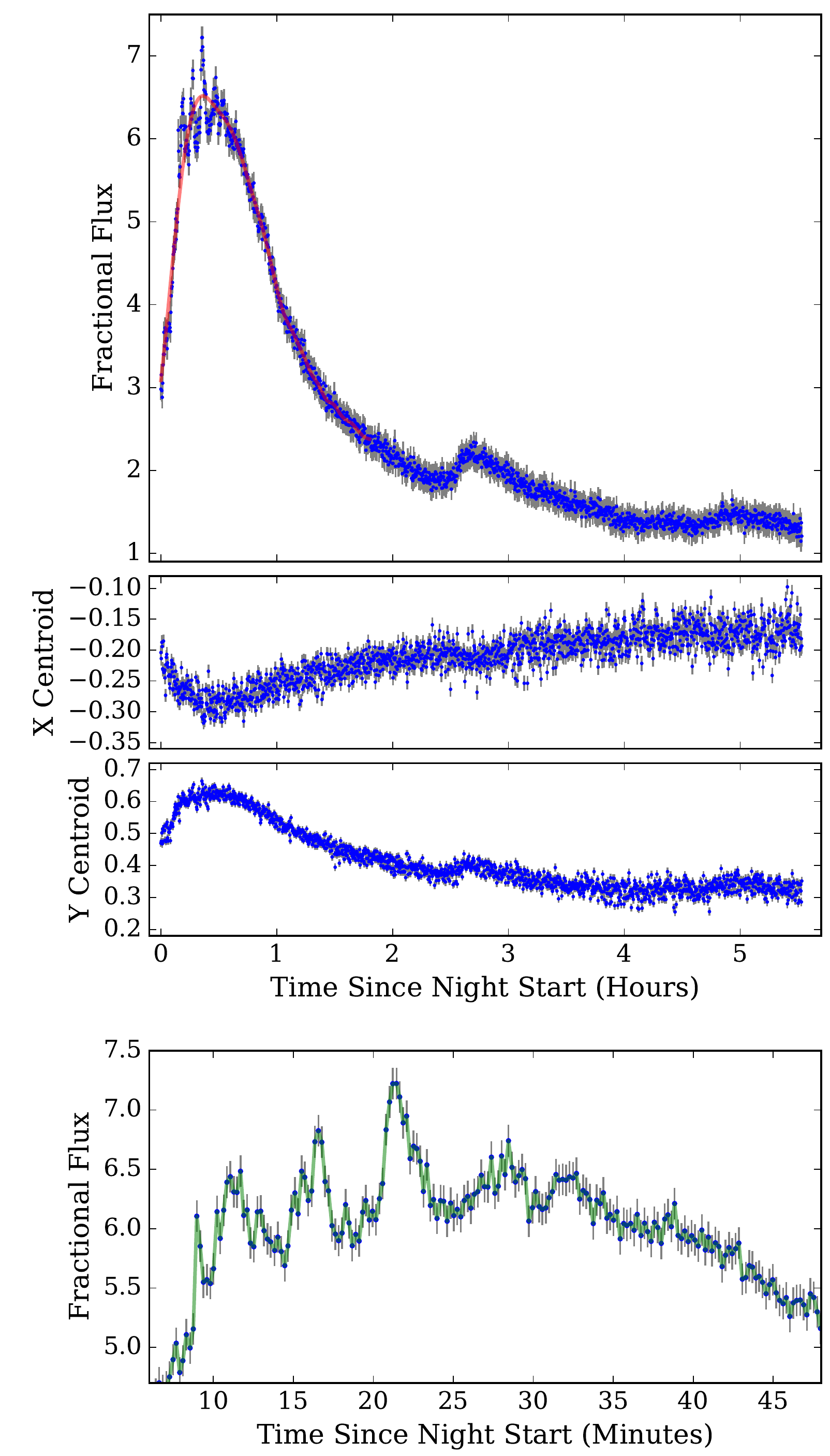}
    \caption{\NGTS\ data from the night of the flare. Top: The flare after the removal of the background source flux, as discussed in Sect.\,\ref{sec:flux_removal}. Not shown are the quiescent nights before and after the flare, which are at an average value of 0 with this normalisation. In red is the trend of the flare from an EMD analysis, discussed in Sect.\,\ref{sec:amp_osc}. Middle: Centroid movement (in pixels) during the night, showing correlation with the flux. X and Y correspond to the movement along the axes specified in Fig.\,\ref{fig:im_fig}. Bottom: Zoom in of the flare peak, in which oscillations are clearly seen. A flux spike, lasting only about 20-30 seconds, is seen at the beginning of the oscillations approximately 8 minutes after the night start. A green interpolating line is shown to aid the eye.}
    \label{fig:flare_fig}
\end{figure}

\begin{figure}
	\includegraphics[width=\columnwidth]{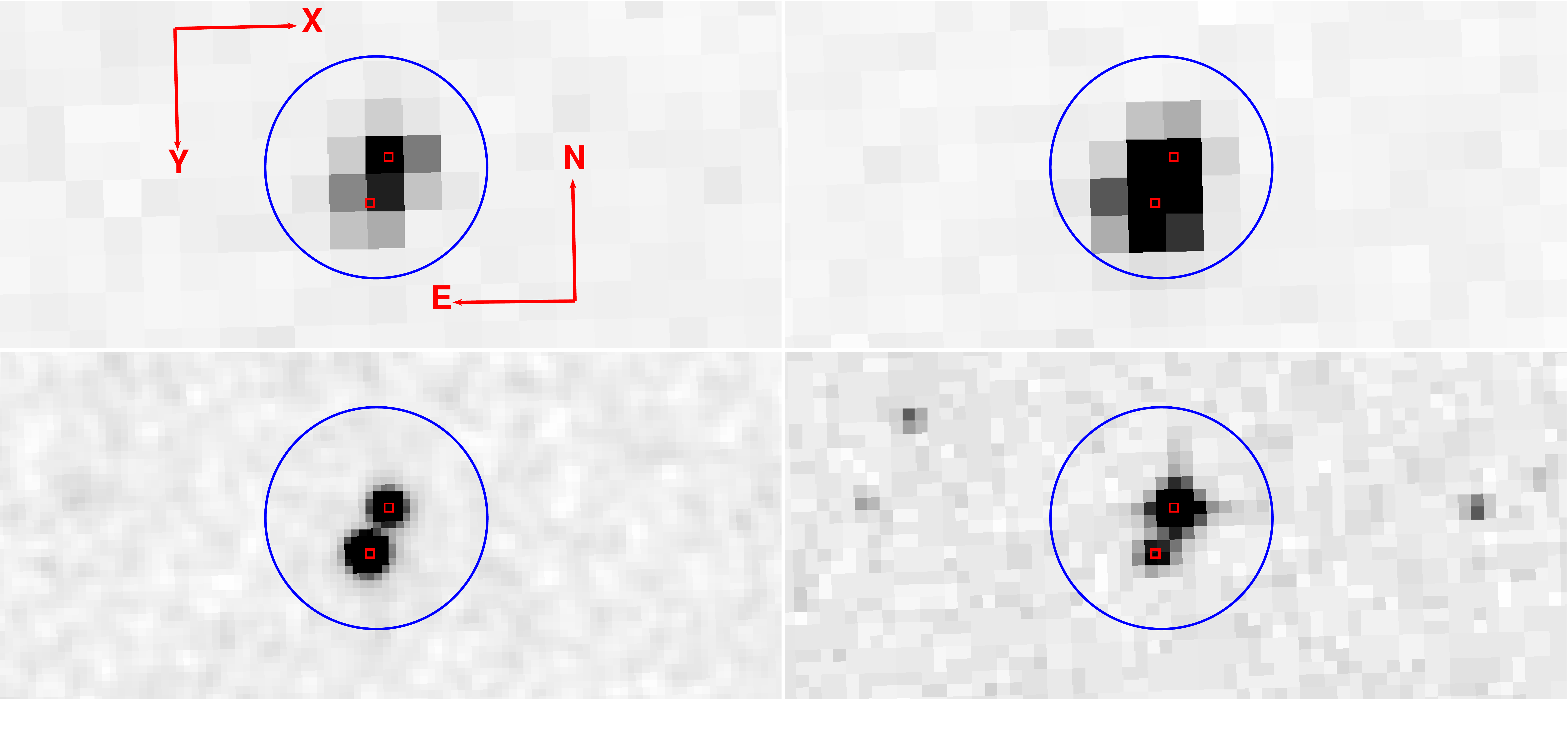}
    \caption{Comparison of NGTS images with 2MASS and SDSS. Top left is our reference image of the source, top right is the image of the source at the peak of the flare. The flaring source is located at the south east. Bottom left is the available 2MASS image and bottom right is from DSS. The blue circle shows our aperture used, red squares are the positions of each individual source from \Gaia. We have also plotted the NGTS image coordinate axes for reference to Fig.\,\ref{fig:flare_fig}.}
    \label{fig:im_fig}
\end{figure}

Following this method we detected a flare from the blended source \longobjname\  (hereafter, \objname). \Gaia\ astrometry reveals that two sources reside within the \NGTS\ aperture, separated by 6.7 arcseconds. The flare was flagged as a night 5\,MAD above the median of the lightcurve and is shown in Fig.\,\ref{fig:flare_fig}. Fig.\,\ref{fig:flare_fig} shows the flare after we have removed the flux contribution from the background star, using the method described in Sect.\,\ref{sec:flux_removal}. The two sources and the \NGTS\ aperture are shown in Fig.\,\ref{fig:im_fig}.

Following the detection of the flare in Fig.\,\ref{fig:flare_fig}, we checked each individual night to find low amplitude flares which may not have been flagged. Through doing this we found three additional lower-amplitude flares towards the end of the season.

\subsection{Centroid Analysis} \label{sec:star_identify}
As two sources are present inside the \NGTS\ aperture, we needed to determine on which star the flare occurred. 
To determine this, we use the centroid shift of the combined stellar image. 
For an isolated source we expect the centre of our aperture $x_{ap}(t)$ to be equal to the centre of flux $x_{flux}(t)$. However, for a blended source there is an offset from the primary source due to the secondary flux contribution. We can define the centroid $\xi(t)$ as the difference in position between the two, or
\begin{equation}
	\xi(t) = x_{flux}(t) - x_{ap}(t).
\end{equation}
During a flaring event we would expect the centroid position to move towards the flaring source. This is similar to the use of centroiding to vet \kepler\ planet candidates \citep[e.g][]{Bryson13}. A more detailed discussion of centroiding in \NGTS\ and its use with blended sources can be found in \citet{Gunther17}. 
\begin{figure}
	\includegraphics[width=\columnwidth]{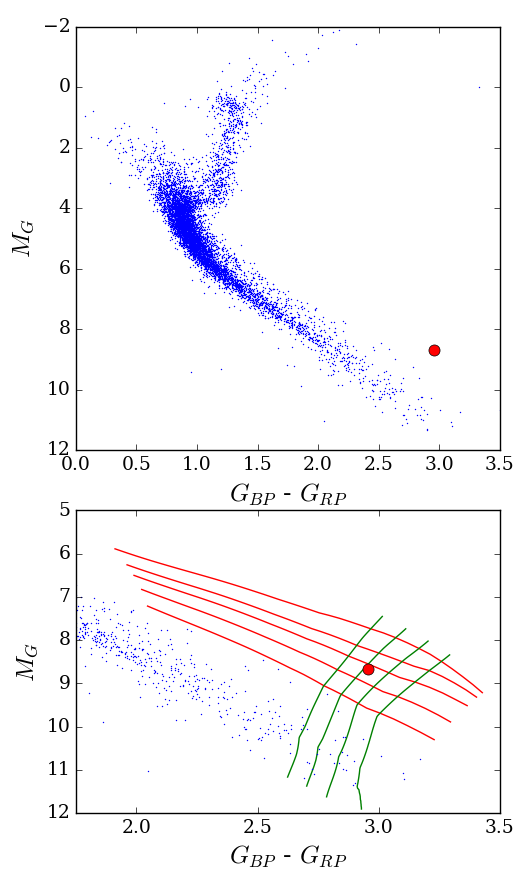}
    \caption{Top: Hertzsprung-Russell diagram of \Gaia\ crossmatched stars in our NGTS field. The red marker indicates the position of \objname\, approximately 1.5 magnitudes above the main sequence. Bottom: The same HR diagram, but now zoomed in and with selected MIST models overlaid. Green lines indicate stellar mass tracks between 0.14 and 0.2~\Msun and red line indicate isochrones of 1, 1.5, 2, 3 and 5 Myr.}
    \label{fig:hr_diagram}
\end{figure}
For this analysis we utilised centroid positions calculated as part of the \NGTS\ data analysis pipeline. Comparison of the centroid position with time for the flare is shown in Fig.\,\ref{fig:flare_fig}. We also compare this centroid movement with the images of the source before and during the flare to confirm the position of the flaring source, identifying the source as \longobjname\ (\objname). This is the south east source in Fig.\,\ref{fig:im_fig} and has previously been identified as \longobjnametwomass. 

\subsection{Stellar Properties}
\subsubsection{Spectral energy distribution} \label{sec:sed_fitting}
\begin{figure}
	\includegraphics[width=\columnwidth]{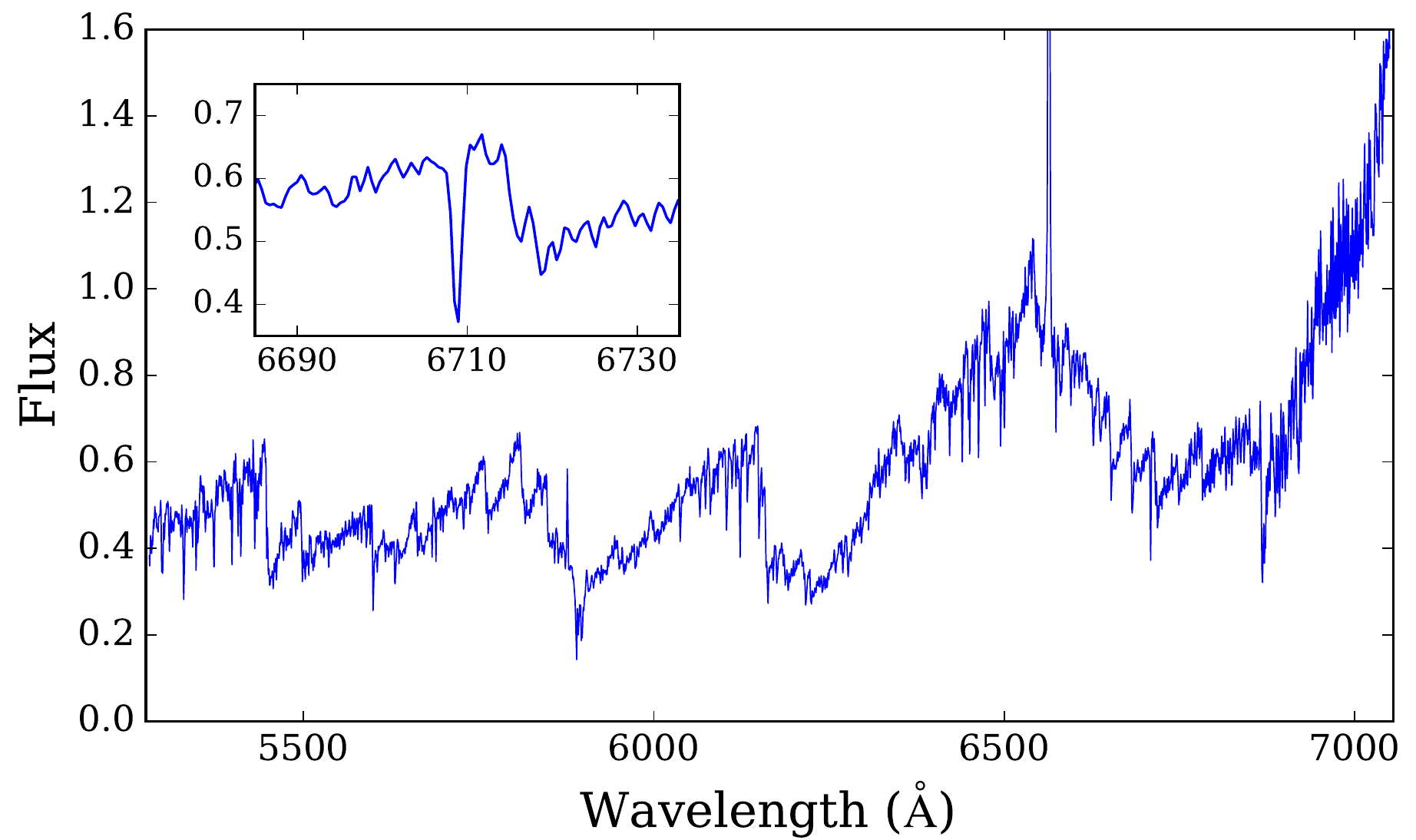}
    \caption{Mean quiescent WiFeS/R7000 spectrum of \objname, normalised at 7000\,\AA. H$\alpha$ continues to a peak flux of 6.5 on this scale. Inset is the \ion{Li}{i} $\lambda$6708 absorption line. }
    \label{fig:QPP_spectrum_plot}
\end{figure}
With this positional information we obtained catalogue magnitudes, making sure to only use catalogues that are able to identify the two sources separately. The results of this matching is shown in Tab.\,\ref{tab:stellar_params}. Using the available 2MASS \citep[][]{2MASS_2006}, \Gaia\ \citep[][]{Gaia_mission,GaiaDR2}, SkyMapper \citep[][]{Wolf18} and WISE \citep[][]{ALLWISE2014} information, we performed spectral energy distribution (SED) fitting using PHOENIX models \citep[][]{Allard16}.  
From this SED fitting and the \Gaia\ DR2 parallax we obtain stellar temperature and radius estimates. 
We find an effective temperature of 3090$\pm$30 K and a radius of 1.04$\pm$0.02~\Rsun\ for the flare star. 

The large radius is not consistent with the expected value for a main sequence M dwarf of this temperature \citep[$\sim$0.2\,\Rsun;][]{Mann15}, suggesting \objname\ is not a single main sequence object. \objname\ is shown on a HR diagram in Fig.\,\ref{fig:hr_diagram}, where it resides around 1.5 magnitudes above stars of the same \gaiabprp\ colour. If it were a binary star we would expect it to reside no more than 0.753 magnitudes above the main sequence \citep[][]{Gaia_HR}, which is not consistent with our observation.

An alternative possibility is that \objname\ is a pre-main sequence object. Interpolating the MESA Isochrones and Stellar Tracks (MIST) v1.1 models \citep[][with updated \Gaia\ DR2 passbands and zeropoints]{Dotter16,Choi16} at the position of \objname\ in Fig.\,\ref{fig:hr_diagram} and assuming zero interstellar reddening, we estimate an age of 2.2~Myr and a mass of 0.18~\Msun. These correspond to a \teff\ of 3180~K and a radius of 0.96~\Rsun, very similar to the values obtained from the SED fit above. We do not identify any IR-excess from our SED fitting on either star, which implies that if \objname\ is a pre-main sequence star it has already lost the majority of any disc that was present during formation. Near- and mid-infrared disc fractions in young clusters and star-forming regions are consistent with median primordial disc lifetimes of 4--5 Myr for solar and later type stars \citep{Bell13,Pecaut16}, suggesting that if \objname\ is as young as 2~Myr, its disc  dissipated earlier than expected, perhaps due to enhanced mass accretion as a result of strong flares \citep[e.g.][]{Orlando11}.
We would expect a low mass star such as \objname\ to be moving almost isothermally along its Hayashi track and thus remain at a similar \gaiabprp\ colour as currently observed. As a further test we compare the 2MASS J,H,K photometry of \objname\ to those from giants and dwarfs in \citet{Bessell88}. Doing this we find that \objname\ resides in the region occupied by dwarfs, supporting our interpretation that \objname\ is a pre-main sequence star as opposed to a subgiant or an unresolved binary system.

From our SED fitting we determine the neighbouring star to be a background G type star. From the \Gaia\ parallax and proper motions it is clear that these two sources are unrelated, with the G type star being much more distant than \objname. The \Gaia\ astrometry for \objname\ are not consistent with membership in any known young moving group, association or open cluster. Although spatially coincident with TW Hya Association members in the Northern reaches of the Scorpius-Centaurus OB association, the $210\pm3$~pc distance to \objname\ means it is likely a young background object or escapee from the region.

\subsubsection{Optical spectroscopy}

To confirm that \objname\ is a pre-main sequence star we acquired six 1800\,s spectra on 2018 July 17 using the Wide Field Spectrograph \citep[WiFeS;][]{Dopita07}
on the ANU 2.3-m telescope at Siding Spring Observatory. The R7000
grating and RT480 dichroic used gave a resolution of $R\approx7000$ over a wavelength range
of 5250--7000\,\AA. Details of the instrument setup and
reduction process are
provided in \citet{Murphy15}. While the absence of strong emission lines other than H$\alpha$ in the first spectrum indicates that the star was likely seen in quiescence, we appear to have serendipitously observed a flare event in subsequent  spectra, with enhanced continuum emission, a rapid increase in the strength and width of the H$\alpha$ and \ion{He}{i} 5876/6678\,\AA\ emission lines and delayed \ion{Na}{d} emission evident. These flare observations will be discussed further in a later work. 

We observed \objname\ twice more in quiescence on July 18 and the average of the three quiescent spectra is plotted in Fig.\,\ref{fig:QPP_spectrum_plot}. \objname\ is clearly of M spectral type, with strong H$\alpha$ emission ($\textrm{EW}=-12\pm1$~\AA) and \ion{Li}{i} $\lambda$6708 absorption  ($\textrm{EW}=610\pm60$~m\AA). The uncertainty in each case is the standard deviation across the three observations. Comparing the spectra to M dwarf templates and radial velocity standards observed each night we estimate a spectral type of M3--3.5 and measure a mean radial velocity of $18.5\pm1.4$ km~s$^{-1}$. Within the limits of WiFeS' modest velocity resolution the star's cross correlation function is consistent with a slowly-rotating ($v\sin i \lesssim 45$~km~s$^{-1}$) single star \citep[see discussion in][]{Murphy15}. We note that the SED and MIST \teff\ values are 200--300\,K cooler than the corresponding pre-main sequence temperature (3360\,K) for an M3 star from the scale of \citet{Pecaut13}. However, their sample of mid-M pre-main sequence stars was dominated by older, higher surface gravity stars from the 10--25 Myr-old $\eta$ Cha, TW Hya, and $\beta$ Pic moving groups, whereas \objname\ may be as young as 2\,Myr old. 

The detection of essentially undepleted lithium in an M3 star is immediate evidence of youth. The MIST models above predict that a 0.18~\Msun\ star will retain its primordial lithium for 30--35~Myr then rapidly deplete the element to zero photospheric abundance by an age of 40~Myr. Any detection of lithium in such a star therefore sets this as an upper age limit. Combining the lithium detection with the position of the star in Fig.\,\ref{fig:hr_diagram}, we are confident that \objname\ is a very young pre-main sequence star.

\subsubsection{X-ray emission} \label{sec:xray}
As a young and active pre-main sequence star, we can expect \objname\ to be a relatively luminous X-ray source.  Fortunately, it was observed serendipitously in X-rays with \XMM\ 
for 27\,ks on 2016 December 20 (ObsID: 0784370101; PI: Loiseau) and inspection of the pipeline-processed images shows that an X-ray source is indeed detected at the position of \objname. The spatial resolution of \XMM\ telescopes is around 5\,arcsec \citep{Jansen01} and the detected X-ray source is clearly centered on the position of the pre-main sequence star, and not the background G star. There is no evidence for significant X-ray emission from this neighbouring object. 

We extracted an X-ray light curve and spectrum for \objname\ from the EPIC-pn camera using the \Gaia\ position and a 20\,arcsec radius aperture. 
The background counts were estimated using a source-free circular region of radius 54\,arcsec located nearby on the same CCD detector. We followed the standard data reduction methods as described in data analysis threads provided with the Science Analysis System\footnote{http://www.cosmos.esa.int/web/xmm-newton/sas} (SAS version 17.0). We found that a background flare had occurred during the final 4.2\,ks of the \XMM\ observation, and we excluded this interval when extracting the X-ray spectrum. 

We inspected the X-ray light curve of \objname\ and found no evidence for significant variability during the \XMM\ observation. This indicates that the observation was free of any large stellar flares and that the X-ray flux can be taken as representative of the quiescent level. 

\begin{figure}
    \includegraphics[width=\columnwidth]{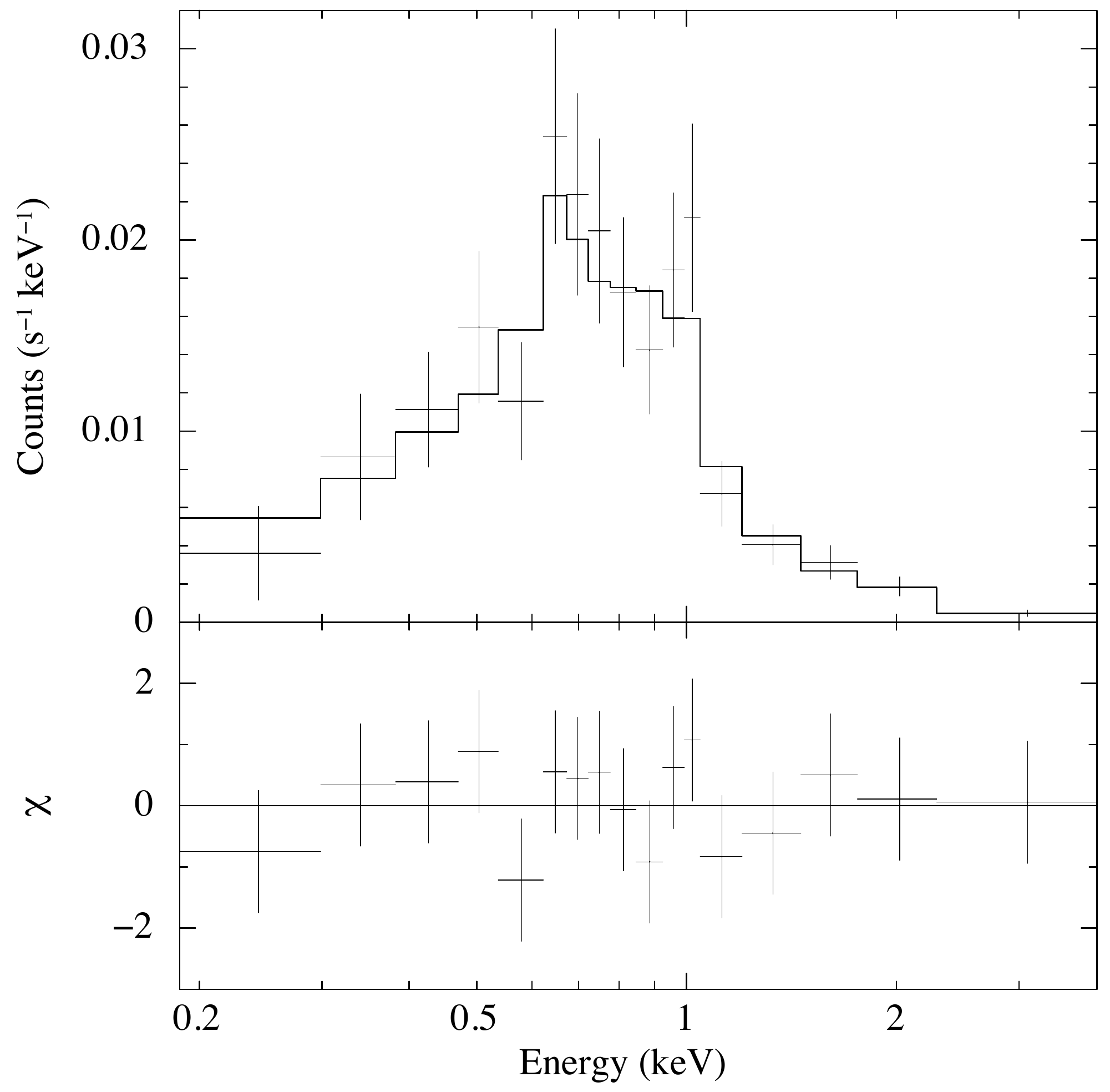}
    \caption{\XMM\ EPIC-pn X-ray spectrum of \objname\ fitted using a two-temperature thermal plasma model, with abundances consistent with the inverse FIP effect, and circumstellar photoelectric absorption. The model and best fitting parameters are described in Sect.\,\ref{sec:xray}.}
    \label{fig:xray}
\end{figure}

The X-ray spectrum of \objname\ is plotted in Fig.\,\ref{fig:xray}. 
We binned the spectrum to a minimum of 20 counts per bin, while ensuring that the spectral resolution would not be oversampled by more than a factor of 3, and we fitted the spectrum using XSPEC\footnote{https://heasarc.gsfc.nasa.gov/xanadu/xspec/} (version 12.10). 

The X-ray spectra of active late-type stars are characterized by optically-thin thermal emission from the corona with temperatures typically in the range $10^6$--$10^7$\,K, and we fitted the EPIC-pn spectrum of \objname\ using the APEC optically-thin thermal plasma model \citep{Smith01}. We found it necessary to include two APEC components (as an approximation to the expected multi-temperature plasma) as well as photoelectric absorption by neutral material \citep[using the TBABS model;][]{Wilms00}. 

We initially set elemental abundances to Solar values \citep{Asplund09}, but we found that the spectral fit was improved significantly around 1\,keV by allowing the Fe and Mg abundances to drop below Solar values (values of 0.03--0.19 Solar). This is consistent with the ``inverse FIP effect'' seen in very active stars and M dwarfs \citep[where FIP refers to the first ionisation potential of the element;][]{Wood10,Wood12,Laming15}. 

Our best fitting model and residuals are also plotted in Fig.\,\ref{fig:xray}. The fitted absorption column density was found to be $N_{\rm H}=1.2\pm^{0.6}_{0.4}\times10^{21}\,\rm cm^{-2}$, which is larger than the total Galactic column density in the direction of \objname\ \citep[$N_{\rm H}=0.65\times10^{21}\,\rm cm^{-2}$;][]{Kalberla05}.
This suggests that the line of sight absorption is dominated by circumstellar material, as might be expected for such a young star. The best fitting temperatures in our model were $0.52\pm^{0.08}_{0.12}$ and $>2.5$\,keV (6 and $>$29\,MK).

The absorbed X-ray flux in the energy range 0.1--2.5\,keV is $3.3\pm^{2.1}_{1.3}\times10^{-14}\,\rm erg\,s^{-1}\,cm^{-2}$, corresponding to an X-ray luminosity of $1.6\pm^{1.0}_{0.6}\times10^{29}\,\rm erg\,s^{-1}$ at a distance of 210\,pc (Table\,\ref{tab:stellar_params}). 
Setting the absorption column density in the model to zero, we can infer the inherent unabsorbed flux to be approximately $6\times10^{-14}\,\rm erg\,s^{-1}\,cm^{-2}$, corresponding to an X-ray luminosity of $3\times10^{29}\,\rm erg\,s^{-1}$. Comparing to the bolometric luminosity of the star (calculated from effective temperature and radius reported in Tab.\,\ref{tab:stellar_params}) gives a ratio $log L_{X}/L_{Bol} = -3.1$. This X-ray activity level is characteristic of saturated X-ray emission, as expected for a young and active star \citep[e.g.][]{de_la_reza04,Stassun04, Santiago10}.   

\subsection{Flux Removal} \label{sec:flux_removal}
To calculate the true flare amplitude and energy we needed to remove the flux contribution from the background G star. To do this we used the PHOENIX spectra from our SED fitting in Sect.\,\ref{sec:sed_fitting} and the NGTS sensitivity curve from \citet{Wheatley18} to estimate the NGTS count rates of the two sources. To account for sky extinction we assumed an atmospheric transmission curve for an airmass of 1.5.

Doing this gave a predicted total flux within 7 per cent of the observed value and the expected ratio of flux from each source. We corrected the predicted total flux to match the observed value, resulting in a quiescent NGTS count rate of \countrate\ ADU/s for the flaring star. We corrected the median level of our lightcurve to this value.

\subsubsection{Stellar Rotation} \label{sec:stellar_rot}
Once we had removed the background flux from the \NGTS\ lightcurve, we investigated the possibility of lightcurve modulation due to stellar rotation from \objname. This was to identify potential starspots and active regions. To do this we used a generalised Lomb-Scargle periodogram from the \textsc{astropy} LombScargle package \citep[][]{Astropy13}. The flares were removed from the lightcurve when calculating the periodogram. We sampled periods between 1 hour and 25 days, requiring 10 samples per peak. This resulted in roughly 60000 periods to sample on \citep[][]{VanderPlas18}. 25 days was chosen as our maximum period to avoid the lunar period. During our analysis we masked suspicious periods due to 1 day aliases. We were unable to identify any convincing period of flux modulation from our analysis.

\section{Results}
\subsection{Flare Amplitude and Energy} \label{sec:flare_amp}
To measure the maximum amplitude of the flare we use the fractional flare amplitude, $\frac{\Delta F}{F}$, which gives the observed flare in units of the quiescent stellar flux. Using the value from Sect.\,\ref{sec:flux_removal} of \countrate\ ADU/s for the median flux level we measured the maximum fractional flare amplitude as \maxampvalue. The quoted error here is larger than that shown in Fig.\,\ref{fig:flare_fig} due to incorporating the error on the quiescent flux of both stars. In Fig.\,\ref{fig:flare_fig} and elsewhere we use the NGTS errors for our analysis.

We calculate the flare energy following the method of \citet{Shibayama13}, assuming the flare emission to be from a blackbody of temperature of 9000$\pm$500 K. Blackbody-like emission from flares has been observed previously on main sequence M dwarfs \citep[e.g.][]{Hawley92} and this method has been used for M dwarf flare calculations elsewhere \citep[e.g][]{Yang17}.
From this, we calculate the bolometric energy of the flare as \flareenergy. As we do not observe the entirety of the flare, this value  acts as a lower limit. It is worth noting that this is roughly 10,000 times greater than the Carrington event energy of $\approx 10^{32}$erg \citep[][]{Carrington_1859, Carrington_Energy}.

We repeated our calculation for the three smaller flares we have identified. Doing this gives energies of \energyone, \energytwo\ and \energythree\ respectively. Consequently, from a total of 484 hours of observations on \objname\ we found four flares above $10^{33}$erg in energy. This gives an estimated occurrence rate of flares above $10^{33}$erg for \objname\ as $72\pm36$ flares per year. 

\subsection{Flare Duration} \label{sec:flare_dur}
To measure the flare duration we make use of two timescales to account for not observing the full extent of the flare.
We first use the e-folding timescale, defined as the time from the flare maximum to 1/e of this value. We measure the e-folding timescale as initially 1.8 hours. However, at around 2.7 hours in Fig.\,\ref{fig:flare_fig} there is a bump which increases the flux back above the 1/e threshold. It is uncertain whether this bump is due to this flaring event or from elsewhere on the flaring star, so we have decided to use our original e-folding timescale of 1.8 hours. From the centroid movement of this bump in Fig.\,\ref{fig:flare_fig} we are confident this bump is not due to the background star. 

Secondly we use the ``scale time'', the duration for which the flare is above half the maximum flux value. This gives a measure of the flare rise without the uncertainties associated with determining where the flare begins. We calculate a scale time of 1.3 hours. We can also put a limit on the minimum and maximum full duration of the flare, using nights before and after. We calculate the absolute minimum and maximum durations as 5.5 and 42.4 hours respectively. 

\subsection{Significance of the Oscillations} \label{sec:signiftest} \label{sec:powspec}
In order to determine whether a statistically significant periodic component is present in the flare light curve, and if so to estimate its period, the Fourier power spectrum during the flare was examined, using the method from \citet{2017A&A...602A..47P}. 

The finite nature of time series data combined with the trends and astrophysical noise seen in flares means that the exact shape of the power spectrum will vary depending on where the start and end times of the flare light curve are defined. Therefore the start and end times that gave the most visible periodic signal in the power spectrum were chosen manually. Flare power spectra tend to have a power that is related to the frequency by a power law, which may be due to trends and/or the presence of red noise, and this power law dependence needs to be carefully accounted for when assessing the significance of a periodic signal. For this part of the analysis, the first step was to fit the power law dependence with a simple model, which was done in log space where the power law appears as a straight line. A broken power law model was used to account for white noise that starts to dominate at higher frequencies \citep{2011A&A...533A..61G}, resulting in a levelling off of the power law, and this model can be written:
\begin{equation}
	\log\left[\hat{\mathcal{P}}(f)\right] = 
		\begin{cases}
			-\alpha\log\left[f\right] + c & \text{if } f < f_{break} \\
			-\left(\alpha - \beta\right)\log\left[f_{break}\right] - \beta\log\left[f\right] + c & \text{if } f > f_{break}\,,
		\end{cases}
\end{equation}
where $f_{break}$ is the frequency at which the power law break occurs, $\alpha$ and $\beta$ are power law indices, and $c$ is a constant. Estimates of the uncertainties of the fitted model were made by performing 10,000 Monte Carlo simulations using the uncertainties of the original light curve data. The following initial parameters were used: $\alpha = 2.0$, $\beta = 0.1$, $c = -0.1$, $\log f_{break} = -1.8$, and these were allowed to vary with a standard deviation equal to 10\% of the parameter values in order to reduce of the possibility of the least-squares fit finding a local minimum rather than the global minimum.

Two approaches were used in order to assess the significance of a peak in the power spectrum corresponding to a periodic signal, which are based on the method described by \citet{2005A&A...431..391V} and account for data uncertainties as well as the power-law dependence of the spectrum. These approaches, which make use of regular and binned power spectra, are given in detail in \citet{2017A&A...602A..47P}.

\begin{figure}
	\includegraphics[width=\columnwidth]{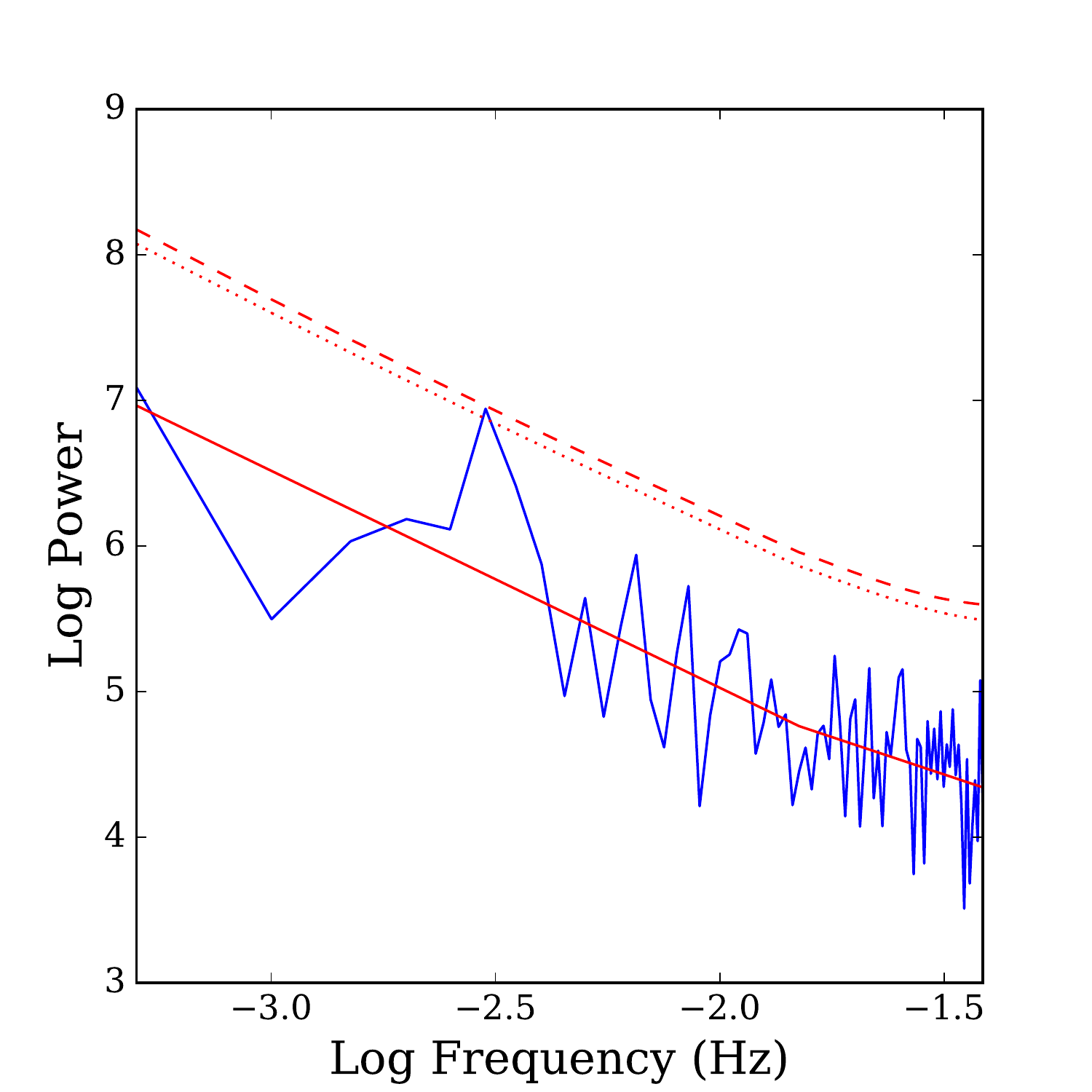}
    \caption{The regular power spectrum for the flare, shown in blue. The solid red line indicates the power law fit, while the two dashed lines are the 95 and 99\% confidence levels respectively. This uses data between HJDs of 2457419.6607291666 and 2457419.6839814815.}
    \label{fig:regpowspec}
\end{figure}

\begin{figure}
	\includegraphics[width=\columnwidth]{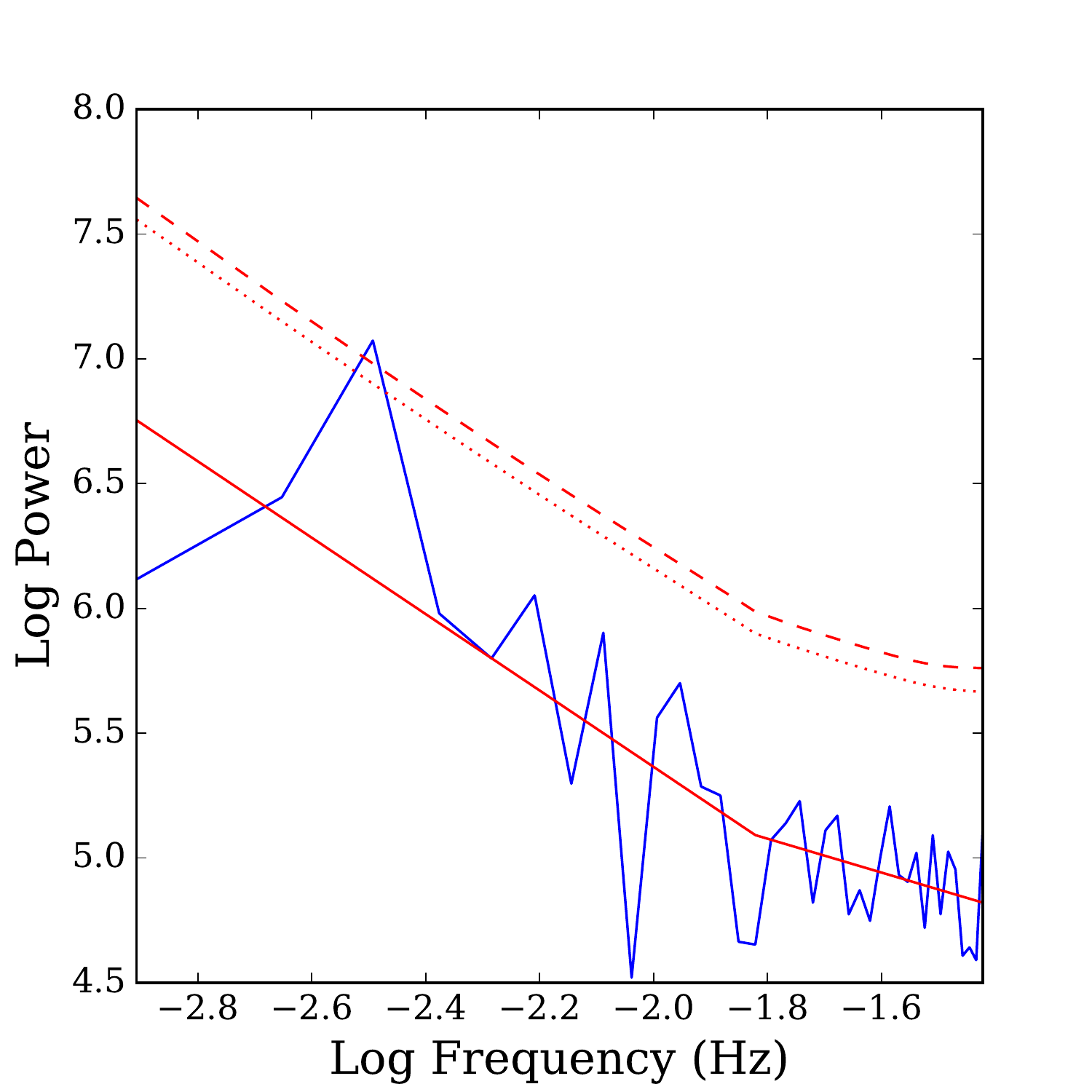}
    \caption{The binned power spectrum for the flare. This uses data between HJDs of 2457419.6607291666 and 2457419.684270833.}
    \label{fig:rebinpowspec}
\end{figure}

From this analysis we find a peak in the regular power spectrum corresponding to a period of $320^{+40}_{-35}$\,s (or $5.2^{+0.7}_{-0.6}$\,min) which reaches the 98.8\% confidence level, as shown by Fig.~\ref{fig:regpowspec}. The uncertainty was taken to be the standard deviation of a Gaussian model fitted to the peak. Because the peak in Fig.~\ref{fig:regpowspec} appears to span more than one frequency bin, if we sum together the powers in every two frequency bins and assess the significance of the same peak in this binned power spectrum, shown by Fig.~\ref{fig:rebinpowspec}, then the peak reaches the 99.8\% confidence level. Hence the periodic component of the flare light curve is highly significant and very unlikely to be the result of noise.


\subsection{Empirical Mode Decomposition} \label{sec:amp_osc}
\begin{figure}
	\includegraphics[width=\columnwidth]{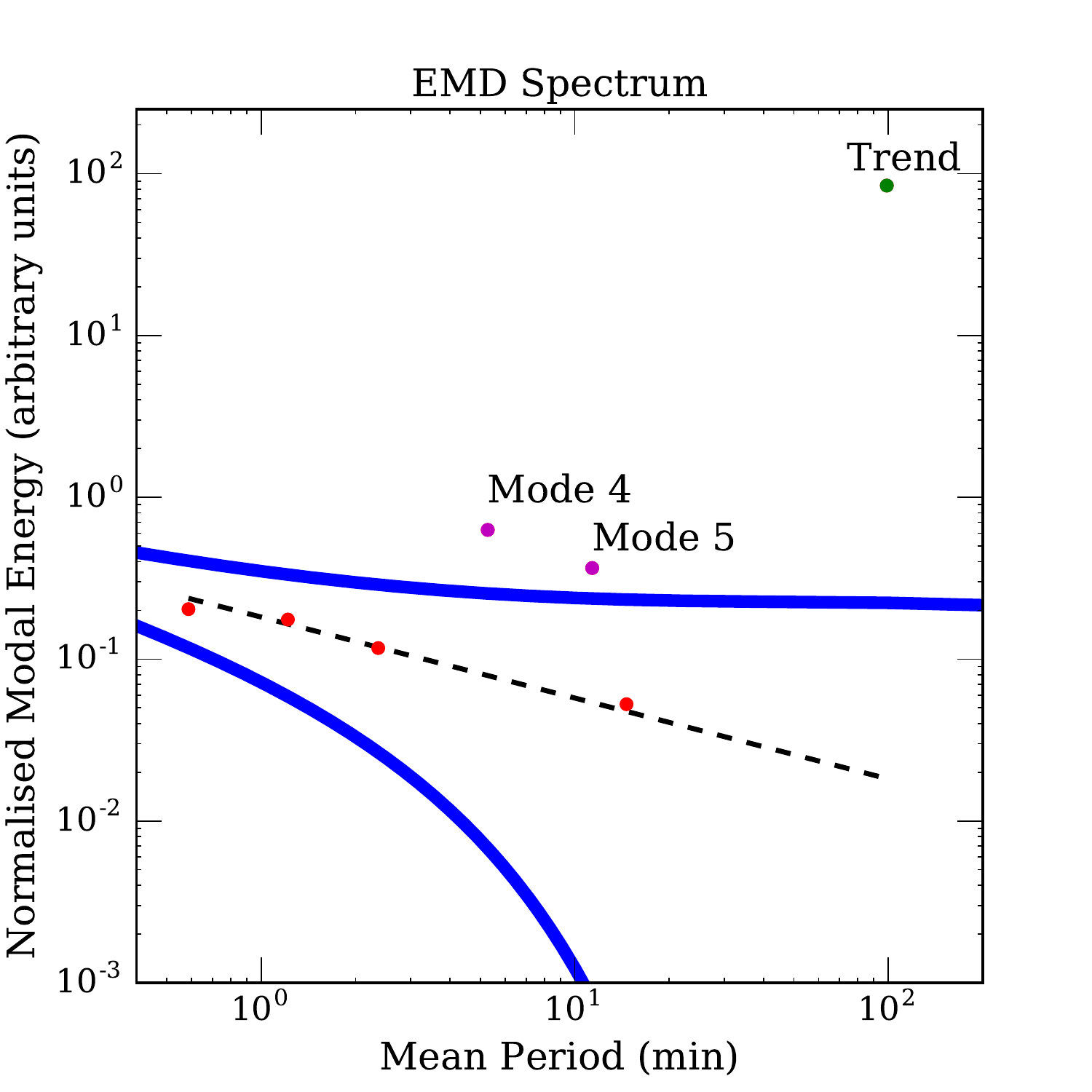}
    \caption{EMD spectrum for the identified modes, the normalised energy $E_m$ versus the mean period $P_m$ for each mode is shown. Modes 4 and 5 are shown with magenta points, the trend is shown in green, all other modes are in red. Plotted in black is the expected behaviour ($E_mP_m^{1-\alpha}=\mathrm{const}$) for noise with $\alpha\approx 0.5$. Plotted in blue are the 99 per cent uncertainty levels, showing modes 4, 5 and the trend to be statistically significant.}
    \label{fig:EMD_spec}
\end{figure}

\begin{figure}
	\includegraphics[width=\columnwidth]{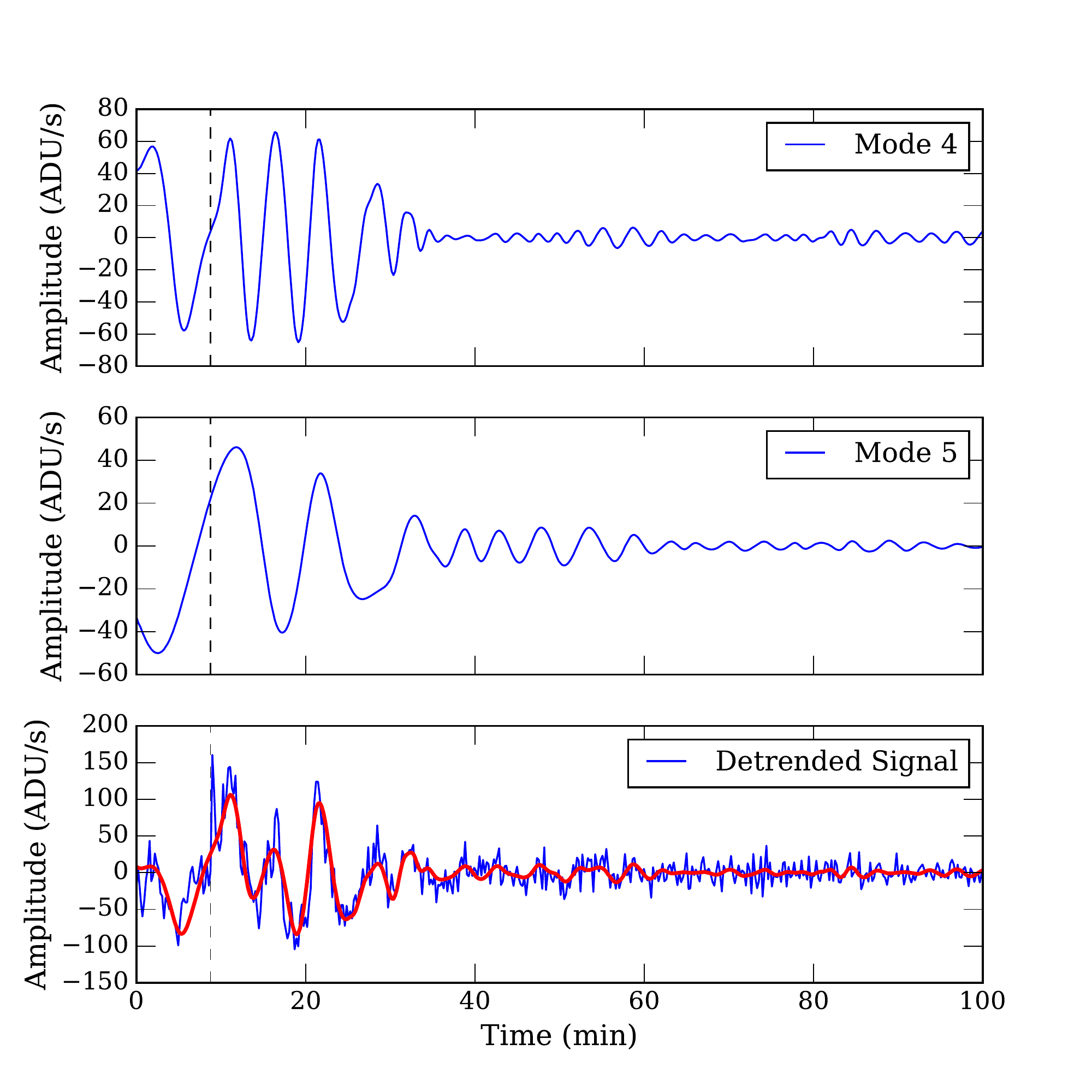}
    \caption{Modes 4 and 5 from the EMD analysis. The bottom row shows them combined (red) over the trend subtracted flare peak. The dashed vertical line indicates the start of a flux pulse, discussed in Sect.\,\ref{sec:discussion}.}
    \label{fig:IMF_Modes}
\end{figure}
An alternative method of investigating the periodicity of QPP signals is to use empirical mode decomposition (EMD) \citep[e.g.][]{Kolotkov15, Cho16}. This method has been utilised previously for solar and stellar QPPs and is used to 
reduce signals into intrinsic mode functions (IMFs), which can be used to describe natural timescales appearing in the original dataset. In particular, the combination of modes can be used to model the longer timescale flare shape. We perform EMD on our flare and obtain seven modes, including the background trend. This trend contains the longest timescale behaviour and is used to fit for the overall flare shape (see Fig.\,\ref{fig:flare_fig} for this trend on the flare peak), as it does not include substructure. 

In order to determine whether these modes are statistically significant we follow the method of \citet{Kolotkov16}, of which we give a brief overview. When performing EMD of coloured noise, the obtained IMFs have the relation $E_{m}P_{m}^{1-\alpha}=$const, where $E_{m}$ and $P_{m}$ are the energy density and the modal period of each IMF and $\alpha$ is the power law index. $\alpha$ is used to describe the colour of the noise present, from the classical definition of the Fourier power spectral density, $S$, of the noise, $S = C/f^{\alpha}$. For example, $\alpha$ = 0 for white noise, 1 for pink noise and 2 for red noise. Using this IMF energy-period relation and assuming all modes follow the same noise power, plotting the energy and period of each mode in logarithmic space and using a linear fit should give $\alpha$. However, any modes with different behaviour, due to some underlying process, will lie off this line. We perform this for our modes, giving $\alpha \approx$ 0.5, shown in Fig.\, \ref{fig:EMD_spec}, suggesting low levels of correlated noise. What can be seen in Fig.\,\ref{fig:EMD_spec} is that modes 4, 5 and the background trend lie off this fitted model, suggesting additional signals are present in these modes which cannot be explained by noise. Using the full method from \citet{Kolotkov16} we calculate 99 per cent significance levels for $\alpha$=0.5 noise, shown by the blue lines in Fig.\,\ref{fig:EMD_spec} for the upper and lower confidence levels. As a result, we see that modes 4 and 5 are statistically significant and proceed to analyse their morphology, regarding the other modes as noise. The trend is also outside this confidence interval, as expected.

For part of the above method we calculated the average modal periods of modes 4 and 5. These are 316 and 682 seconds respectively. As another test we also follow the method of \citet{Kolotkov15} and use a Hilbert transform to calculate their instantaneous frequencies and periods. Constraining our region to the same area used in Sect.\,\ref{sec:powspec} we obtain median instantaneous periods of 338 and 625 seconds for modes 4 and 5 respectively. To obtain an estimate of the period error we use the mean and the standard deviation of these periods, giving periods of 327$\pm$11 and 654$\pm$30 seconds for modes 4 and 5 respectively. 

\subsubsection{QPP Signal and Amplitude}
Fig.\,\ref{fig:IMF_Modes} shows modes 4 and 5, along with their combination and the trend-subtracted flare. We can see that the combination of both modes is able to reproduce the observed QPPs. We note here that the periodicity of mode 4 appears to begin after a 20 second duration spike in flux, approximately 8 minutes into the night, whereas mode 5 is present from the start of the night. This spike in flux is also clearly visible in Fig.\,\ref{fig:flare_fig}. We can use the trend in conjunction with the QPP signal to also determine the fractional flux amplitude, \oscfrac, of the oscillations. Here we have denoted $F_{tr}$ as the trend flux. Doing this and ignoring the flux spike we obtain the fractional flux of the combined oscillations \oscfrac=0.1.

\subsubsection{Wavelet Analysis}
\begin{figure}
	\includegraphics[width=\columnwidth]{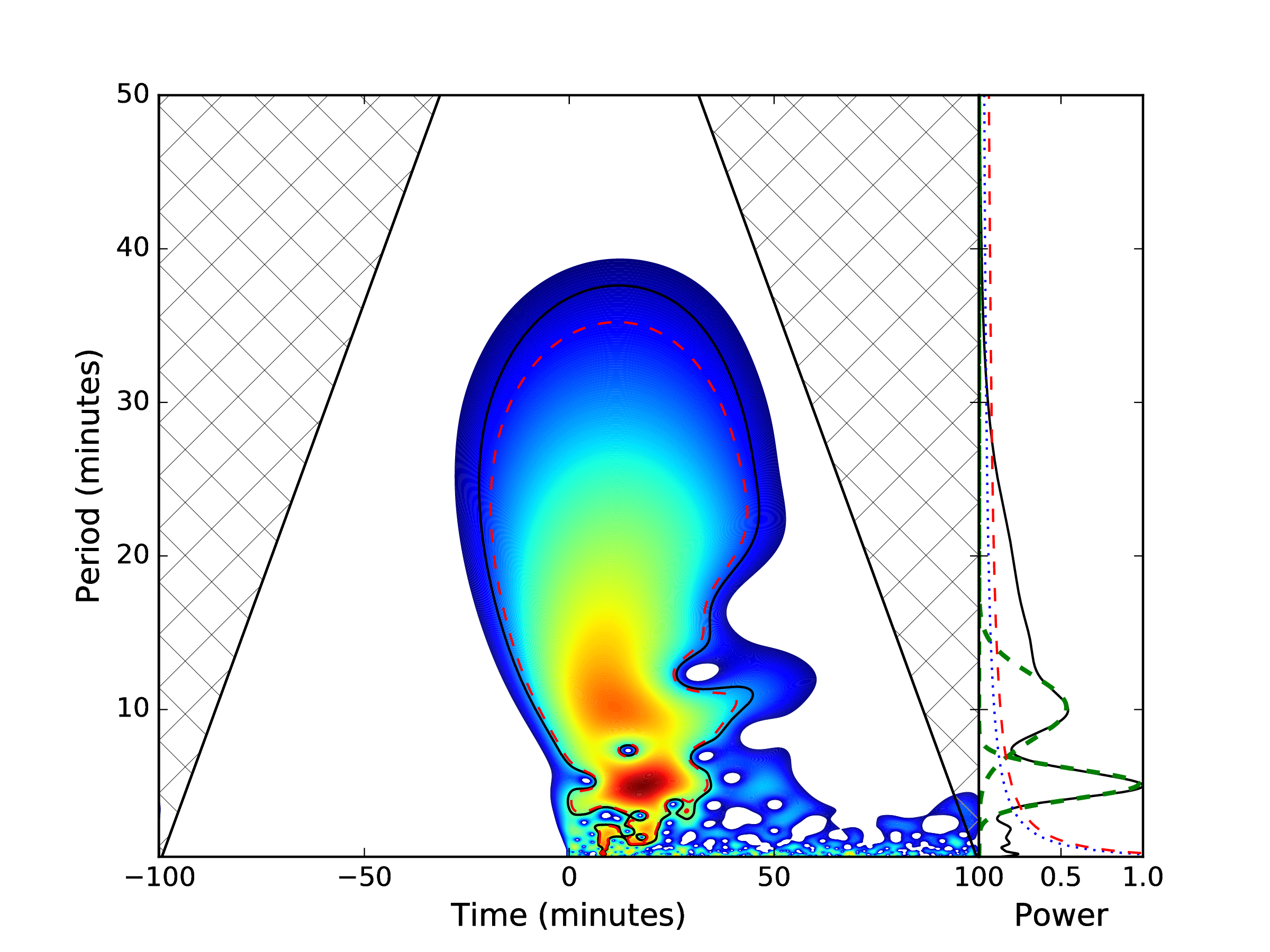}
    \caption{Wavelet analysis of the flare, showing two peaks at 309.2 and 608.5 seconds respectively. Red depicts the periods with greatest power. The black wavelet boundary is the 99 per cent confidence level and the red line is the modified confidence level, based on \citet{Auchere16}. The right hand panel shows the peaks of the two components, with Gaussian fits in green.}
    \label{fig:wavelet_plot}
\end{figure}
In order to help confirm the periodicities from our previous two methods, we have performed a wavelet analysis.
This was done incorporating the method of \citet{Auchere16}, which takes into account the total number of degrees of freedom of the wavelet spectrum when assessing the significance of identified peaks in power. To perform this analysis we used a detrended lightcurve, obtained by dividing our lightcurve by the overall flare trend identified through our EMD analysis. The results of our wavelet analysis are shown in Fig.\,\ref{fig:wavelet_plot}. From Fig.\,\ref{fig:wavelet_plot} we can see two peaks in power, at 309 and 609 seconds. These agree with the period identified from our power spectrum analysis and the two periods identified with the EMD analysis, adding weight to there being two periods present. The ratio between these periods suggests that they may be harmonics of the same MHD mode, something which we discuss in Sect.\,\ref{sec:MHD_modes}. The 309 second period appears to show a slight offset from the 609 second peak, something that is also seen in our EMD analysis and is also discussed in more depth in Sect.\,\ref{sec:MHD_modes}.

\begin{table}
	\centering%
	\label{tab:flare_params}
	\begin{tabular}{|l|c|}
    \hline
    Property & Value \tabularnewline \hline
   	Energy (erg) & >\flareenergynoerg\  \tabularnewline
    Trend Amplitude (per cent) & 650  \tabularnewline
    Peak Flare Amplitude (per cent) & 720  \tabularnewline
    Min Duration (Hours) & 5.5 \tabularnewline
    Max Duration (Hours) & 42.4 \tabularnewline
	e-folding duration (Hours) & 1.8-2.6 \tabularnewline
    Scale time (Hours) & 1.3 \tabularnewline
    QPP Period, Fourier (s) & $320^{+40}_{-35}$ 
    \tabularnewline
    QPP Period, EMD - modal  (s) & 316 (Mode 4), 682 (Mode 5) \tabularnewline
    QPP Period, Wavelet (s) & 309 (Mode 4), 609 (Mode 5) \tabularnewline
    QPP Amplitude (per cent) & 10 \tabularnewline
	\hline
	\end{tabular}
    \caption{Properties of the superflare detected from \objname.}
\end{table}
\section{Discussion} \label{sec:discussion}
We have detected a large white-light stellar flare from the pre-main sequence M star \longobjname. This star is extremely young with an estimated age of 2.2 Myr and undepleted lithium in the stellar spectrum. From SED fitting and optical spectroscopy we estimate an spectral type of M3-3.5. The flare displays multi-mode quasi-periodic pulsations in the peak. 
We have calculated the energy of this flare as \flareenergy\ and the maximum amplitude as \maxampvalue. Through multiple methods we have identified the periods of oscillations as approximately 320 and 660 seconds, with an oscillation amplitude (\oscfrac) of 0.1. The 320 second oscillation mode appears to begin after a 20-30 second spike in flux during the flare rise. The observation of this spike and the resolution at which the 320 second mode is seen is a testament to the high cadence of \NGTS\ compared to the \Kepler\ and \TESS\ short cadence modes, which would not obtain such detail. 

\subsection{Flare Energy} \label{sec:flare_size}
In Sect.\,\ref{sec:flare_amp} we calculated the lower limit of the flare energy as \flareenergy. This energy is greater than all M dwarf flares observed with \Kepler\ by \citet{Yang17} and is comparable to that emitted by the highest energy G star superflares \citep[e.g.][]{Shibayama13,Wu15}.
This value sits within the range of bolometric flare energies calculated for 3 Myr stars in NGC2264 by \citet{Flaccomio}, using targeted \corot\ observations. It is also similar to the energies of flares observed from long cadence K2 observations of the 1 Myr brown dwarf CFHT-BD-Tau 4 by \citet{Paudel18}. 

As mentioned in Sect.\,\ref{sec:flare_amp}, due to us not observing the entirety of the flare this energy is a lower limit. 
Previous works that have observed only a portion of long duration flare events have been able to estimate the full energy through using empirical flare models to obtain the full flare shape \citep[][]{Schmidt16}. These models typically assume the flare is a ``simple'' event, i.e. with no significant substructure \citep[e.g.][]{Davenport14,Jackman18}. However, ``complex'' flares such as the one observed do not strictly follow such models and have a range of morphologies (e.g. multiple peaks). Consequently, we have decided against applying an empirical flare model to estimate the full energy.

When calculating the bolometric energy, it was assumed that the flare could be modelled as a 9000K blackbody. 
This method assumes that the spectrum of the flare is constant for the total duration. Previous observations have shown this may not be the case \citep[e.g.][]{Kowalski13} and has resulted in other works resorting to alternative methods. One commonly used method is to calculate the equivalent duration of the flare and multiply this by the quiescent luminosity of the star in the instrument bandpass \citep[e.g.][]{Hawley14,Davenport16,Vida17}. For our observed flare we have decided not to use such a method. Our reasoning for this is that using the quiescent luminosity in such a way assumes the flare emits with the same spectrum as the star, which for M stars is not the case - particularly over a wide bandpass such as \NGTS. By using a 9000K blackbody we expect to remain close to an expected flare spectrum.

\subsection{Occurrence Rate of Flares} \label{sec:occ_rates}
In Sect.\,\ref{sec:flare_amp} we estimated the occurrence rate for flares above $10^{33}$erg as $72\pm36$ per year for \objname. For the flare in Fig.\,\ref{fig:flare_fig}, a simple scaling of a single detection in 484 hours gives an estimated occurrence rate for flares above $10^{36}$erg for \objname\ alone as 18 per year. A more reliable estimate can be made by assuming stellar flares on \objname\ occur with a power law distribution of flare energies \citep[e.g][]{Lacy76}. By normalising to our lower energy occurrence rate we can estimate the distribution at higher energies.

From X-ray observations of the Orion Nebula Cluster by \citet[][]{Caramazza07}, the power law index of X-ray flares low mass stars has been measured as $2.2\pm0.2$. However, \citet{Flaccomio} have shown from simultaneous optical and X-ray observations that X-ray flares do not always have an optical counterpart. From their analysis, between 19 and 31 per cent of X-ray flares lack an optical counterpart. Consequently, as our observation is of a white-light flare we have decided to make use of power law indices obtained from white-light measurements. 
\jj{The power law index for white-light flares on active M stars has been measured for both individual stars \citep[e.g.][]{Moffett74, Lacy76, Ramsay15} and groups \citep[e.g.][]{Hilton11}. These measurements have shown that active M stars can have a range of power law indices. 
Due to this scatter we have chosen a range of power law indices for predicting the flare rate. We have chosen limits of 1.53 and 2.01, taken from \citet{Hawley14} and \citet{Hilton11} for observations of active M stars.}
We note that the upper limit here is still similar to that from \citet{Caramazza07}. With these values for the power law index and our lower energy occurrence rate, we reestimate our occurrence rates for flares above $10^{36}$ erg from \objname\ as between $3\pm1.5$ per year and $2\pm1$ per decade.

While this is a wide range of values, it provides a more robust estimate of the occurrence rate of high energy flares from \objname. However, it has previously been noted that the occurrence rate drops off at the highest energies, giving a broken power law \citep[e.g.][]{Davenport16}. As this break energy may vary from star to star, we have not incorporated this into our estimations. However, we do note that if our flare is in the ``drop off'' regime then this would make it even rarer than our estimations above. This highlights not only the rarity of this event, but the need for long-duration wide-field surveys such as \NGTS\ to find them.

\subsection{Flare Amplitude}
We can compare this flare to previous white light observations of QPP bearing stellar flares and see that it is the largest observed for its period timescale (between 5 and 15 minutes) and one of the largest detected overall. The maximum change in \NGTS\ magnitude is $\Delta m_{NGTS} = 2.28$ for $\frac{\Delta F}{F}$=\maxampvalue. Of similar oscillation period were the flares from II Peg and EV Lac observed by \citet{Mathioudakis03} and \citet{Zhilyaev00} respectively. These both had observations in the U band, with flare amplitudes of 0.85 and $\approx$ 2.5. As stellar flares are blue in colour (particularly during the impulsive phase), we would expect increased amplitudes in the U band compared to the \NGTS\ bandpass \citep[e.g. as for \kepler][]{Hawley14}. Using our SED fit from Sect.\,\ref{sec:sed_fitting} with the assumed 9000K flare blackbody (using a Sloan SDSS U band filter\footnote{\url{http://svo2.cab.inta-csic.es/svo/theory/fps3/index.php?id=SLOAN/SDSS.u\&\&mode=browse\&gname=SLOAN\&gname2=SDSS\#filter}}) we estimate $\Delta m_{U} = 6$ for the observed flare, equivalent in amplitude to the 32 minute period oscillating megaflare on YZ CMi \citep[][]{Kowalski10,Anfinogentov13}. Consequently, we believe this flare to be one of the largest observed in white-light showing QPPs for both its period timescale and possibly overall.

\subsection{Formation and Habitability of M Star Exoplanet Systems}
Flares and associated Coronal Mass Ejections (CMEs) from pre-main sequence stars could have an important role in planetary formation and habitability.  For example, these transient events have been proposed as a possible mechanism for the formation of chondrules found in meteorites \citep[][]{Feigelson10}. The flash-melting of these rocks requires a transient heat source, which could be provided either by direct absorption of flare XUV irradiation \citep[][]{Shu01}, or through a flare associated shock wave \citep[][]{Nakamoto05}.  

Flares and CMEs have also been linked to high stellar mass-loss rates by \citet{Osten15} for young stars and for those with debris discs, CMEs have been highlighted as a possible cause for observed IR variability. This is due to the removal of IR-emitting dust by CMEs on the timescales of days \citep[][]{Osten13}. For protoplanetary discs large X-ray flares could, along with altering the structure and ionisation fraction of the disc \citep[e.g.][]{Ilgner06}, cause increased mass transfer from the disc onto the star. This is through perturbations from flare loops linking the star and inner disc \citep[][]{Orlando11} which result in bursts of accretion. 

Once planets are formed, they will also be subject to irradiation from flares and CMEs. The transient increase of the stellar wind through CMEs may negatively influence the formation of a planetary dynamo \citep[][]{Heyner12}, resulting in a weakened planetary magnetic field. The planetary magnetic field is one of the main defences against the detrimental effects of CMEs, which can compress the planetary magnetosphere and expose the atmosphere to erosion \citep[e.g.][]{Kay16}. Along with this, X-ray and UV irradiation from the flare itself may cause intense planetary ozone depletion \citep[][]{Segura10}, altered atmospheric chemical abundances \citep[][]{Venot16} and potential damage to the DNA of surface organisms \citep[e.g.][]{Lingam17}. 

M stars have been noted for emitting less steady-state NUV radiation compared to earlier spectral types. This could limit possible UV-sensitive prebiotic chemistry \citep[e.g.][]{Ranjan17}, calling into question whether or not life could appear in these systems. The highest energy flares such as the one observed have been suggested as a possible way of delivering required levels of UV irradiation to kick-start such prebiotic reactions \citep[][]{Buccino07,Ranjan17,Rimmer18}. Due to the rarity of these high energy events, studies of their occurrence rates (as in Sect.\,\ref{sec:occ_rates}) with surveys such as \NGTS\ are vital in helping to determine whether prebiotic chemistry could take place or whether more negative effects limit habitability.

\subsection{MHD Modes} \label{sec:MHD_modes}
Using the fourth and fifth modes generated from the EMD method outlined in Sect.\,\ref{sec:amp_osc} we found that the QPPs could be reproduced. We calculated the average modal periods as 316 and 682 seconds, and their median instantaneous periods as 338 and 625 seconds. 

These instantaneous periods were obtained by using the region of data used in Figure \ref{fig:regpowspec}, from the flare peak. We note that this still includes a small section in the flare tail where the oscillations are indistinguishable from the noise in the data. Excluding this and constraining the Hilbert spectrum region further results in instantaneous periods of 330 and 660 seconds for modes 4 and 5. The period for mode 4 is in agreement with the period obtained from Sect.\,\ref{sec:powspec} and this brings the period of mode 5 into better agreement with its average modal period. Therefore the period ratio for modes 4 and 5 is approximately two, as would be expected if these modes were fundamental and secondary harmonics, but we note that even for harmonics this period ratio can deviate from two due to the dispersive nature of some MHD waves and the local geometry \citep[e.g.][]{Inglis09}.

We note also that the appearance of modes 4 and 5 are similar to those of modes 4 and 5 from \citet{Kolotkov15}, which were oscillation periods of 45 and 100 seconds and were associated with potentially being harmonics of the MHD kink mode. 
The period ratio of modes 4 and 5 suggests that they are both from the same MHD process, possibly a standing slow magnetoacoustic wave. In this mode, excitations of both the fundamental and second harmonic can be produced, with their relative amplitudes being dependent on the location of the flare trigger. Specifically, pulses closer to the flare loop footpoint will excite the fundamental mode, whereas those closer to the apex will excite the second harmonic \citep[][]{Selwa05}. \citet{Selwa05} also found that pulse triggers in locations other than the apex and footpoint can result in the excitation of a packet of standing waves with different modes, with the lowest-frequency two modes having the greatest contribution.

If this were the case, we may expect mode 4 to have an appearance similar to mode 5, namely that of an exponentially decaying sinusoid. However, the appearance of mode 4 is more similar to that of previously observed wave trains \citep[e.g.][]{Nakariakov04}, which occur due to fast magneto-acoustic waves propagating in the plasma non-uniformity. These wave trains are highly dispersive in nature and can be created via an impulsive driving pulse or perturbation \citep[][]{Roberts84}. We observe a spike in flux during the flare rise, immediately after which the observed periodicity of mode 4 begins.
This flux spike lasts for approximately 20-30 seconds in total, meaning it would not be identified in \Kepler\ or \TESS\ short cadence observations. The amplitude of this spike is greater than the detrended signal, making it possible that this is a driving pulse which triggers the appearance of mode 4. A similar example has been observed in the Sun by \citet{Nistic14}, where a pulse from a single source set off a quasi-periodic wave train. We do not see a similar change in period from mode 5 in Fig.\,\ref{fig:IMF_Modes}. 

This behaviour is also seen in Fig.\,\ref{fig:wavelet_plot} from our wavelet analysis, where the shorter period peak is offset from the longer period. Consequently we propose that this is a broadband driving pulse which results in the excitation of a quasi-periodic wave train \citep[][]{Nakariakov09,Nakariakov16}, resulting in mode 4. This wave train could then be a combination of fast magnetoacoustic harmonics.

\subsubsection{Mode 5}
As mentioned previously, mode 5 does not share a similar appearance to mode 4 and seems generally unaffected by the observed flux spike. Instead it has the appearance of a decaying mode with a damping time of $\sim$20 minutes. Comparing this damping time to the oscillation period of $\sim$660 seconds, we find it is in agreement with the upper limit of the relation found by \citet{Ofman02}. This relation is for the oscillation period and damping timescale of hot flare loops observed by the SUMER instrument aboard \textit{SOHO} \citep[][]{Wang11}.

To test the hypothesis that mode 5 is indeed due to a standing slow mode, we can compare the period ratio between our two modes. If we assume the coronal loop to be represented by a magnetic cylinder of radius $a$ and length $L$, in the low $\beta$ limit (where magnetic pressure dominates) we can write the period of mode 4 as
\begin{equation} \label{eq:fast_period}
P_1 = \frac{2\pi a}{j_{0,1}v_A}\bigg(1-\frac{\rho_{e}}{\rho_o}\bigg)
\end{equation}
where $v_A$ is the \Alfven\ speed within the loop, $\rho_{e}$ and $\rho_o$ are the external and internal plasma densities and $j_{0,1}$ is the first zero of the Bessel function $J_{0}(z)$ \citep[][]{Roberts84}. For $\rho_e << \rho_o$, the fast mode is highly dispersive. Consequently, the bracketed term reduces to unity and we can rewrite equation \ref{eq:fast_period} as 
\begin{equation} \label{eq:new_fast_period}
P_1 \approx \frac{2.62a}{v_A}
\end{equation}
For a standing slow mode the period, $P_2$, is given by \citet[][]{Roberts84} as
\begin{equation}
P_2 \approx \frac{2L}{c_s}
\end{equation}
where $c_s$ is the internal sound speed of the cylinder. From our observations, $\frac{P_2}{P_1} \approx 2$, so
\begin{equation}
\frac{P_2}{P_1} = \frac{2L}{c_s} \cdot \frac{v_A }{2.62a}
\end{equation}
which approximates to $\frac{v_A L}{c_s a}$. Hence, $\frac{L}{a} \approx 2\frac{c_s}{v_A}$. The \Alfven\ speed is typically greater than the sound speed, resulting in $L<a$, namely the length of the cylinder is less than its radius. This is an unrealistic expectation for coronal loops and is not supported by observations from solar flares, making it unlikely mode 5 is a standing slow mode.
An alternative option is that mode 5 is instead a standing kink mode with a decaying amplitude. The period of a standing kink mode is given by \citet[][]{Edwin83} as
\begin{equation}
P_2 \approx \frac{2L}{c_k}
\end{equation}
where $c_k$ is the internal kink speed and in this low $\beta$ regime is approximately $\sqrt[]{2}v_A$ \citep[e.g.][]{Nakariakov01}. From this, we can write the period ratio $\frac{P_2}{P_1}$ as
\begin{equation}
\frac{P_2}{P_1} = \frac{2L}{c_k} \cdot \frac{v_A}{2.62a}
\end{equation}
which approximates to $\frac{L}{2a} = 2$, or $\frac{L}{a} = 4$, which is closer to what we would expect from a coronal cylinder \citep[e.g.][]{Van11}. 
Standing kink oscillations 
associated with flaring events have been observed previously on the Sun \citep[e.g.][]{Ofman08,Zimovets15,Goddard16} and have previously been suggested as the cause for QPPs in some observed stellar flares \citep[e.g.][]{Anfinogentov13}. Consequently, we propose that a plausible explanation for mode 5 is a standing kink mode which was triggered at the start of the flare.
\subsection{Seismology of mode 4}
The generation of impulsive fast waves can be used to investigate the behaviour of the oscillating region. To do this, we first rearrange equation \ref{eq:new_fast_period}, to obtain the ratio between $a$ and $v_A$. We can then substitute in our observed fast mode period of 320s to determine the ratio as $a/v_{A}$ = 123 s.

If we assume the cylinder radius of $1\times10^{10}$cm, similar to X-ray flares previously observed on pre-main sequence stars \citep[e.g.][]{Santiago10,Favata05}, we estimate the \Alfven\ speed as $8\times10^{7}\textrm{cm\,s}^{-1}$. This estimated value is similar to those calculated for QPPs from main sequence M4 stars \citep[e.g.][]{Zaitsev04,Mitra05,Anfinogentov13}, suggesting that our assumed cylinder radius value is a sensible one. Following this, if we use an aspect ratio of $\simeq$ 0.25 \citep[e.g.][]{Mathioudakis03} then the loop length is $\simeq 0.55\textrm{R}_{*}$. Assuming the loop is semicircular we also estimate the height as $\simeq 0.18\textrm{R}_{*}$, similar to the average loop height for X-ray detected flares from pre-main sequence stars studied by \citet{Flaccomio}.

This ratio of loop length to stellar radius, while too large for the Sun and similar spectral types, is consistent with previous inferred measurements of loop lengths on main-sequence M dwarfs \citep[which can go up to 2$R_{*}$ in length,][]{Mullan06} and is well within loop lengths for pre-main sequence stars \citep[e.g.][]{Favata05,Johnstone12}. Indeed, for stars with a measured NIR excess \citet{Favata05} measured loop lengths up to tens of times the stellar radii, with the flaring loops connecting the star and disc. For stars in their sample with no measurable NIR excess (such as we find for \objname) \citet{Favata05} found more compact loop lengths, similar to what we find for \objname. These smaller loops were suggested to be anchored into the photosphere only.

\subsection{Flare Decay} \label{sec:flare_decay}
We also note the flare exhibits a ``bump'' at about 2.6 hours after the start of the night and another at 4.9 hours. Relative to the flare peak, they are 2.3 and 4.6 hours afterwards. Bumps have been observed in \kepler\ stellar flares by \citet{Balona15} who argued that they cannot be due to simultaneous independent flaring events, nor due to forced global oscillations. One possibility is that while these bumps are not from independent flares, they are instead due to sympathetic flaring. Sympathetic flaring occurs when the primary flare triggers a successive flare, due to a physical connection \citep[][]{Moon2002}. Such behaviour would result in the observed flux increases in the flare tail, making it a possible cause. 

\begin{table*}
	\centering
	\begin{tabular}{lccccr} 
		\hline
		Parameter & & & Value (Flaring) & Value (Companion) & Reference\\
		\hline
        & & & Position & &\\
        \hline
        RA ICRS & & &184.9153111262$\degree$ & 184.9144202 & 1\\
        Dec ICRS & & &-35.9338316576$\degree$ &  -35.9321149 & 1 \\
        \hline
        & & & Photometric & &\\
        \hline
        SkyMapper $u$& & & N/A & 15.930 $\pm$ 0.009 & 2\\
        SkyMapper $v$ & & & N/A & 15.523 $\pm$ 0.008 & 2\\
        SkyMapper $g$ & & & 17.060 $\pm$ 0.009 & 14.521 $\pm$ 0.003 & 2\\
        SkyMapper $r$ & & & 16.125 $\pm$ 0.011 & 14.227 $\pm$ 0.005 & 2\\
        SkyMapper $i$ & & & 14.379 $\pm$ 0.014 & 14.063 $\pm$ 0.008 & 2\\
        SkyMapper $z$ & & & 13.651 $\pm$ 0.004 & 14.019 $\pm$ 0.009 & 2\\
        \Gaia\ $G$ & & & 15.286 $\pm$ 0.001 & 14.237 $\pm$ 0.001 & 1\\
        \Gaia\ $BP$ & & & 16.972 $\pm$ 0.009 & 14.576 $\pm$ 0.001 & 1\\
        \Gaia\ $RP$ & & & 14.017 $\pm$ 0.003 & 13.738 $\pm$ 0.001 & 1\\
        $J$ & & & 12.086 $\pm$ 0.024 & 13.162 $\pm$ 0.030 & 3\\
		$H$ & & & 11.520 $\pm$ 0.025 & 12.846 $\pm$ 0.041 &3\\
		$K_{s}$ & & & 11.160 $\pm$ 0.021 & 12.842 $\pm$ 0.034 & 3\\
        $W1$ & & & 11.000 $\pm$ 0.024 & 12.575 $\pm$ 0.049 & 4\\
		$W2$ & & & 10.792 $\pm$ 0.021 & 12.594 $\pm$ 0.052 & 4\\
		$W3$ & & & 10.778 $\pm$ 0.071 & 12.867 $\pm$ 0.463 & 4\\
		$W4$ & & & 8.704 & 8.680 & 3\\
		\hline
        & & & Kinematic & & \\
        \hline
        $\mu_{RA}$ [mas $yr^{-1}$] & & & -19.177 $\pm$ 0.108 & -2.653 $\pm$ 0.108 & 1\\
        $\mu_{DEC}$ [mas $yr^{-1}$] & & & -8.490 $\pm$ 0.083 & -4.644 $\pm$ 0.037 & 1\\
        Parallax [mas] & & & 4.733 $\pm$ 0.072 & 1.098 $\pm$ 0.032 & 1\\
        Distance [pc] & & & 210 $\pm$ 3 & 887 $\pm$ 26 & \\
        \hline
        & & & SED Fit & & \\
        \hline
        $T_{eff}$ [K] & & & 3090 $\pm$ 30 & 5610 $\pm$ 30 & \\
        $R$ [$R_{\sun}$] & & & 1.05 $\pm$ 0.02 & 1.22 $\pm$ 0.03 & \\
        \hline
        & & & X-ray emission & & \\
        \hline
        $L_X$ [$\rm erg\,s^{-1}$] & & & $3\times10^{29}$ & & \\
        $\log L_{\rm X}/L_{\rm Bol}$ & & & $-3.1$ & & \\
        \hline
	\end{tabular}
    \caption{Parameters for our sources. References are 1. \Gaia\ \citet{GaiaDR2} 2. SkyMapper DR1.1 \citet{Wolf18} 3. 2MASS \citet{2MASS_2006}. 4. WISE \citet{ALLWISE2014}. For distances we have used the values from \citet{BailerJones18}. SkyMapper magnitudes are AB magnitudes.}
    \label{tab:stellar_params}
\end{table*}

\section{Conclusions}
In this work we have detected a high energy stellar flare from the 2 Myr old pre-main sequence M star \longobjname\ with \NGTS. This flare has a minimum energy of \flareenergy, making it one of the largest energy M star flares observed. In the peak of this high energy flare we have detected statistically significant quasi-periodic pulsations with an oscillation amplitude \oscfrac\ of 0.1. We have applied techniques typically used for analysis of solar flare QPPs to determine that the pulsations were formed of two distinct modes. The periods of these modes are approximately 320 and 660 seconds. With a measured amplitude of $\Delta m_{NGTS} = 2.28$ and an estimated amplitude of $\Delta m_{U} = 6$ we believe this is one of the largest white-light stellar flares to show QPPs of this timescale, if not QPPs in general.

Investigating these modes further, we have identified that the shorter period mode appeared after a high amplitude spike in flux during the flare rise. This spike lasted for 30 seconds and was only resolvable due to the high cadence of \NGTS. We postulate that the short period mode is a highly dispersive fast mode excited by the observed flux spike, similar to events seen in the Sun. We hypothesize the longer period mode is a kink mode excited at the flare start, however we cannot categorically rule out other proposed methods for the QPP excitation \citep[][]{2018SSRv..214...45M}. 

We have also detected three more lower energy flares in our data. Using these to estimate the flare occurrence rate, we find the high energy flare to be a rare event, with a possible occurrence rate of between $3\pm1.5$ per year and $2\pm1$ per decade, depending on the power law index of the assumed flare distribution. We use this to stress the importance of wide field, long timescale surveys such as \NGTS\ in finding these high energy events, which must also have high cadence in order to characterise oscillation modes. This is not only for constraining their occurrence rates, but in helping to determine their role in the formation and habitability of Earth-like exoplanets around M-type stars. 

\section*{Acknowledgements}
This research is based on data collected under the NGTS project at the ESO La Silla Paranal Observatory. The NGTS facility is funded by a consortium of institutes consisting of 
the University of Warwick,
the University of Leicester,
Queen's University Belfast,
the University of Geneva,
the Deutsches Zentrum f\" ur Luft- und Raumfahrt e.V. (DLR; under the `Gro\ss investition GI-NGTS'),
the University of Cambridge, together with the UK Science and Technology Facilities Council (STFC; project reference ST/M001962/1). 
JAGJ is supported by STFC PhD studentship 1763096.
PJW and RGW are supported by STFC consolidated grant ST/P000495/1.
AMB acknowledges the support of the Institute of Advanced Study, University of Warwick and is also supported by STFC consolidated grant ST/P000320/1.
MNG is supported by STFC award reference 1490409 as well as the Isaac Newton Studentship.
CEP acknowledges support from the European
Research Council under the SeismoSun Research Project No. 321141. 
JSJ acknowledges support by Fondecyt grant 1161218 and partial support by CATA-Basal (PB06, CONICYT). DYK acknowledges the support by the STFC consolidated grant ST/P000320/1.
GMK is supported by the Royal Society as a Royal Society University Research Fellow.
We also acknowledge and thank the ISSI team led by AMB for useful discussions. 

This publication makes use of data products from the Two Micron All Sky Survey, which is a joint project of the University of Massachusetts and the Infrared Processing and Analysis Center/California Institute of Technology, funded by the National Aeronautics and Space Administration and the National Science Foundation.
This publication makes use of data products from the Wide-field Infrared Survey Explorer, which is a joint project of the University of California, Los Angeles, and the Jet Propulsion Laboratory/California Institute of Technology, funded by the National Aeronautics and Space Administration.
The national facility capability for SkyMapper has been funded through ARC LIEF grant LE130100104 from the Australian Research Council, awarded to the University of Sydney, the Australian National University, Swinburne University of Technology, the University of Queensland, the University of Western Australia, the University of Melbourne, Curtin University of Technology, Monash University and the Australian Astronomical Observatory. SkyMapper is owned and operated by The Australian National University's Research School of Astronomy and Astrophysics. The survey data were processed and provided by the SkyMapper Team at ANU. The SkyMapper node of the All-Sky Virtual Observatory (ASVO) is hosted at the National Computational Infrastructure (NCI). Development and support the SkyMapper node of the ASVO has been funded in part by Astronomy Australia Limited (AAL) and the Australian Government through the Commonwealth's Education Investment Fund (EIF) and National Collaborative Research Infrastructure Strategy (NCRIS), particularly the National eResearch Collaboration Tools and Resources (NeCTAR) and the Australian National Data Service Projects (ANDS). This work has made use of data from the European Space Agency (ESA) mission
{\it Gaia} (\url{https://www.cosmos.esa.int/gaia}), processed by the {\it Gaia}
Data Processing and Analysis Consortium (DPAC,
\url{https://www.cosmos.esa.int/web/gaia/dpac/consortium}). Funding for the DPAC
has been provided by national institutions, in particular the institutions
participating in the {\it Gaia} Multilateral Agreement.



\bibliographystyle{mnras}
\bibliography{references} 




\bsp	
\label{lastpage}
\end{document}